\documentclass[reprint,
 amsmath,amssymb,
 aps,
]{revtex4-2}

\usepackage{graphicx}% Include figure files
\usepackage{dcolumn}% Align table columns on decimal point
\usepackage{bm}% bold math
\newcommand{\bra}[1]{\left\langle{#1}\right\rvert}
\newcommand{\ket}[1]{\left\lvert{#1}\right\rangle}

\newcommand{\vect}[1]{\mathbf{#1}}
\newcommand{\Eq}[1]{Eq.~\eqref{#1}}
\newcommand{\I}{\mathrm{i}}

\begin{document}

%\preprint{APS/123-QED}

\title{Three-boson scattering hypervolume for a nonzero orbital angular momentum
%Three-body scattering hypervolume of $L=2$ system in three dimensions
}% Force line breaks with \\

\author{Pui In Ip and Shina Tan}
\email[shinatan@pku.edu.cn] \\

\affiliation{%
 International Center for Quantum Materials,Peking University, Beijing 100871, China\\
}%

\date{\today}% It is always \today, today,
             %  but any date may be explicitly specified

\begin{abstract}
We analyze the zero energy collision of three identical bosons in the same internal state with total orbital angular momentum $L=2$,
assuming short range interactions. By solving the Schrödinger equation asymptotically, we derive two expansions of the wave function when three bosons are far apart or a pair of bosons and the third boson are far apart. The scattering hypervolume $D$ is defined for this collision.
Unlike the scattering hypervolume defined by one of us in 2008 \cite{tan2008three}, whose dimension is length to the fourth power, the dimension of $D$ studied in the present paper is length to the eighth power. We then derive the expression of $D$ when the interaction potentials are weak, using the Born's expansion. We also calculate the energy shift of such three bosons with three different momenta $\hbar \vect k_{1}$, $\hbar\vect k_{2}$ and $\hbar\vect k_{3}$ in a large periodic box. The obtained energy shift depends on $D^{(0)}/\Omega^{2}$ and $D/\Omega^{2}$, where $D^{(0)}$ is the three-body scattering hypervolume defined in Ref.~\cite{tan2008three} for the three-body $L=0$ collision and $\Omega$ is the volume of the periodic box. We also calculate the contribution of $D$ to the three-body T-matrix element for low-energy collisions.
We then calculate the shift of the energy and the three-body recombination rate due to $D^{(0)}$ and $D$ in the dilute homogeneous Bose gas. The contribution to the three-body recombination rate constant from $D$ is proportional to $T^2$ if the temperature $T$ is much larger than the quantum degeneracy temperature but still much lower than the temperature scale at which the thermal de Broglie wave length becomes comparable to the physical range of interaction.
\end{abstract}

%\keywords{Suggested keywords}%Use showkeys class option if keyword
                              %display desired
\maketitle

%\tableofcontents

\section{\label{sec:level1}Introduction}
When studying the low-energy effective interactions of particles with short-range interaction potentials,
one can focus on those partial waves with small orbital angular momenta. %\cite{bloch2008many,Luciuk2016}[remark]. 
The two-body systems with total orbital angular momenta equal to zero, one, two, etc, have been studied in the literature. %\cite{tan2008three,bijit2024,andrade2023,tan2005}[remark].
%In this paper, we focus on the system in which total orbital angular momentum quantum number $L=2$.

In order to illustrate the physical property of low-energy two-body scattering process, we only need to determine a small number of physical parameters, such as the $s$-wave scattering length $a_s$, the $p$-wave scattering volume $a_p$, and the $d$-wave scattering hypervolume $a_d$. Many of these parameters can be extracted from the wave functions
for zero-energy collisions. For two-body collisions at zero energy in the center-of-mass frame, there are an infinite number of wave functions which behave differently at large interparticle distances. Let $\vect s$ be the spatial vector from one particle to the other particle. The wave function at zero energy approaches a harmonic polynomial of $\vect s$
plus small corrections at $s\to\infty$. In the three-dimensional space, there are $(2l+1)$ linearly independent homogeneous harmonic polynomials of degree $l$, and they all correspond to the orbital angular momentum $l$.
So we can study zero energy wave functions which behave like $s^l$ at large $s$.
The most important of these wave functions for purposes of understanding low-energy effective interactions is the one which grows at the slowest rate at large $s$ (because in such quantum states, the probability of finding
the two particles within the range of interaction is maximized),
and it is the $s$-wave collision wave function. If one considers two identical bosons in the same internal state,
the wave function must be symmetric and hence $l$ must be even, and so the most important zero energy wave function
beyond the $s$-wave is the $d$-wave collision wave function which grows like $s^2$ at large $s$.

When we study the low-energy effective interactions of three particles with short-range interactions, we can follow a similar route:
we can study the zero-energy wave functions in the center-of-mass frame first (meaning that the wave functions we are studying should be \emph{independent of} the three-body center-of-mass coordinates). There are an infinite number of such wave functions, and the most important of them are those which grow at the slowest rates when the three particles are far apart. If we consider three identical bosons in the same internal state, the most important such zero-energy wave function is the one which approaches a constant (can be set to $1$) when the three bosons are far apart, and this wave function has been studied in Ref.~\cite{tan2008three} and some subsequent papers \cite{shinazhu2017,mestrom2019}.
In the present paper, however, we would like to study the most important three-boson zero-energy wave functions
in the center-of-mass frame
beyond the one studied in the aforementioned papers. Considering the symmetry of the three-boson wave function under the interchange of any two bosons, one can easily show that these second-most-important wave functions
grow as quadratic harmonic polynomials of the nine Cartesian coordinates of the three bosons when the three bosons
are far apart, and there are five linearly independent such quadratic harmonic polynomials,
and they all correspond to orbital angular momentum quantum number $L=2$. They can be distinguished according to the magnetic quantum number $M$ (which can take values $-2,-1,0,+1,+2$), where $M\hbar$ is the projection of the orbital angular momentum along the $z$ axis.

%\cite{hammer2010causality,zhai2021,uedabec}.

 In analogy with the two-body scattering, in which the scattering length (or volume or hypervolume) determines the effective strength of low-energy interactions, there is a physical parameter called the three-body scattering hypervolume $D$ which characterizes the three-body scattering \cite{tan2008three,shina2021three}. $D$ was first defined in Ref.~\cite{tan2008three} and it determines the effective strength of three-body interactions at small collision energies. $D$ can affect the energy of the dilute Bose-Einstein condensate (BEC). The imaginary part of $D$ is related to the three-body recombination. In general, $D$ can be calculated numerically \cite{shinazhu2017}. People have computed $D$ for different types of interactions, Gaussian \cite{shinazhu2017}, hard sphere \cite{tan2008three}, square well \cite{mestrom2019}, and Lennard-Jones potential \cite{mestrom2020}. 
 The definition of $D$ depends on the dimension of the three-body system \cite{shina2021three,shinaliang2024,shinazhang2025,shinawang20212}. 
 In the fermionic systems \cite{shinawang2022,shinawang2023} and a two-dimensional bosonic system \cite{shinaliang2024}, the three-body parameters have been defined.

In this paper, we consider three identical bosons in the same internal state with short-range interactions (i.e., we assume that the interaction disappears when the interparticle distance is greater than a range $r_e$)
colliding with zero energy in the center-of-mass frame, with a wave function which grows quadratically when the three pairwise distances are large;
the orbital angular momentum quantum number of the three-body state we study in this paper is $L=2$, and the magnetic quantum number is $M$
satisfying $-L\le M\le L$.
We derive two asymptotic expansions of the wave function of the three-body state, one of which corresponds to the limit in which the three pairwise distance are all large and it is called the 111-expansion. The other one is called the 21-expansion which is applicable when two bosons are held at a fixed distance while the remaining one is far away from the  two. The scattering hypervolume $D$ appears in these two expansions. 
We then study several physical applications of the hypervolume $D$.

In section~\ref{sec:2}, we derive the 111-expansion and 21-expansion of the three-body wave function. In the 111-expansion, we expand the wave function to the order of $B^{-6}$, where $B$ is the hyperradius and is defined in \Eq{eq:hyperradius} below. In the 21-expansion, we expand the wave function to the order of $R^{-6}$, where $R$ is the (large) distance between the center-of-mass of the two bosons held at any fixed distance and the remaining boson which is far away. The three-body scattering hypervolume $D$ appears at the order of $B^{-6}$ in the 111-expansion,
and appears at the order of $R^{-6}$ in the 21-expansion.
Unlike the result obtained in Ref.~\cite{tan2008three}, in which the dimension of the three-body scattering hypervolume for three bosons colliding with zero orbital angular momentum is length to the fourth power, the dimension of $D$ derived in this paper is length to the eighth power. If the two-body interaction is sufficiently attractive to support at least one bound state, $D$ will in general have a nonzero imaginary part \cite{shinazhu2017}, and at the end of this section, we derive a relation between the imaginary part of $D$ and the probability amplitudes for the production
of bound pairs flying apart from the remaining free boson.

In section~\ref{sec:3}, we consider three bosons with \emph{weak} interaction potentials
and study the Born series for the wave function. Comparing the Born series with the 111-expansion, we obtain an approximate formula of $D$ for weak interaction potentials.

In section~\ref{sec:4}, we derive an approximate formula for the energy shift of three bosons in a large periodic box caused by three-body scattering hypervolumes.

In section~\ref{sec:5}, we derive the T-matrix element of the three-body scattering in the absence of two-body potentials at low-energy. We show that the T-matrix element is linearly dependent on the three-body scattering hypervolumes.

In section~\ref{sec:6}, we derive an approximate effective second-quantized Hamiltonian for a Bose gas with nonzero three-body scattering hypervolumes and negligible two-body interactions.

In section~\ref{sec:7}, we study the energy shift of a dilute Bose gas due to the adiabatic introduction of $D^{(0)}$ and $D$ in the thermodynamic limit. We obtain the energy shifts below and above the critical temperature $T_{c}$ of the Bose-Einstein condensation.

In section~\ref{sec:8}, we derive an approximate formula for the three-body recombination rate constant $L_3$ of the dilute Bose gas in terms of the imaginary parts of the three-body scattering hypervolumes, and study its temperature dependence. %When the temperature $T$ is much larger than the quantum degeneracy temperature $T_d\equiv\hbar^2n^{2/3}/(2M_Bk_B)$ (where $\hbar$ is Planck's constant over $2\pi$, $n$ is the number density of the Bose gas, $M_B$ is the mass of each boson, and $k_B$ is the Boltzmann constant), the contribution to $L_3$ from $D$ defined in this paper is proportional to $T^{2}$.

\section{Asymptotic expansions of the three-body wave function \label{sec:2}}
In this section we consider a system of three identical bosons with mass $M_B$ each, colliding at zero energy and zero total linear momentum. We assume that the interactions between these bosons have a finite range and depend solely on the interparticle distances. The Schrödinger equation can be written as
\begin{equation}
\begin{split}
&\biggl[-\frac{\hbar^2}{2 M_B}(\nabla^{2}_{1}+\nabla^{2}_{2}+\nabla^{2}_{3})+V(s_{1})+V(s_{2})+V(s_{3})\\&
+V_\text{3-body}(s_1,s_2,s_3)\biggr]\Psi(\vect{r}_1,\vect{r}_2,\vect{r}_3)=0 ,
\label{eq:sc}
\end{split}
\end{equation}
where $\vect{r}_i$ is the position vector of the $i$th boson, and
\begin{subequations}\label{s}
\begin{align}
    \vect s_1&\equiv\vect r_2-\vect r_3,\\
    \vect s_2&\equiv\vect r_3-\vect r_1,\\
    \vect s_3&\equiv\vect r_1-\vect r_2.
    \label{eq:si}
\end{align}
\end{subequations}
$V(s)$ is the two-body potential, and $V_\text{3-body}(s_1,s_2,s_3)$ is the three-body potential.
We assume that $V(s)=0$ at $s>r_e$ and $V_\text{3-body}(s_1,s_2,s_3)=0$ when $s_1$ or $s_2$ or $s_3$ is greater than $r_e$, where $r_e$ is the range of the interaction.
Since the total linear momentum of the three bosons is zero, the wave function is transitionally invariant:
\begin{equation}\label{translationalinvariance}
   \Psi(\vect r_1+\delta \mathbf{r},\vect r_2+\delta \mathbf{r},\vect r_3+\delta \mathbf{r})=\Psi(\vect r_1,\vect r_2,\vect r_3),
\end{equation}
where $\delta \mathbf{r}$ is an arbitrary vector. We define the Jacobi coordinates $\vect s_i$ and $\vect R_i$ \cite{Glockle1983,Stumpf1974}, where $\vect s_i$ was defined in Eqs.~\eqref{s}, and
\begin{subequations}
\begin{align}
\vect R_1\equiv\vect r_1-\frac{\vect r_2+\vect r_3}{2},\\
\vect R_2\equiv\vect r_2-\frac{\vect r_3+\vect r_1}{2},\\
\vect R_3\equiv\vect r_3-\frac{\vect r_1+\vect r_2}{2}.
\end{align}
\end{subequations}
We also define the hyperradius $B$ and hyperangles $\theta_1$, $\theta_2$, and $\theta_3$:
\begin{align}
    B&=\sqrt{R_i^2 + \frac{3}{4}s_i^2}=\sqrt{\frac{s_1^2+s_2^2+s_3^2}{2}},\label{eq:hyperradius} \\
    \theta_i &= \arctan \left(\frac{2 R_i}{\sqrt{3} s_i}\right).
\end{align}
$R_i$ and $s_i$  can be expressed in terms of $B$ and $\theta_i$:
\begin{eqnarray}
    R_i = B \sin\theta_i,\; s_i = \frac{2}{\sqrt{3}} B \cos\theta_i.
\end{eqnarray}

Equations~\eqref{eq:sc} and \eqref{translationalinvariance} do \emph{not} uniquely define the three-boson zero-energy scattering state.
To fix the definition of $\Psi(\vect r_1,\vect r_2,\vect r_3)$, we need to specify its behavior at large hyperradii.
When $s_1$, $s_2$, and $s_3$ are all large, $\Psi(\vect r_1,\vect r_2,\vect r_3)$ satisfies the Laplace equation in nine-dimensional space: $(\nabla_1^2+\nabla_2^2+\nabla_3^2)\Psi=0$.
So the leading order term in the 111-expansion of $\Psi$ should be a harmonic polynomial of $\vect r_1,\vect r_2,\vect r_3$.
The lowest order harmonic polynomial is $1$, and the associated three-body wave function has been studied in Ref.~\cite{tan2008three}.
The next order harmonic polynomials consistant with Bose statistics and translational invariance are quadratic, and there are five linearly independent
such polynomials, and they all correspond to the orbital angular momentum quantum number $L=2$. They can be distinguished using the magnetic quantum number $M$, satisfying $-L\le M\le L$.
So we consider a wave function satisfying
\begin{equation}
\Psi=T^{(2)}(\vect r_1,\vect r_2,\vect r_3)+O(B^1)
\end{equation}
when $s_1$, $s_2$, and $s_3$ go to infinity simultaneously, where
\begin{equation}\label{T2}
T^{(2)}(\vect r_1,\vect r_2,\vect r_3)\equiv4\sqrt{\frac{2\pi}{15}}\sum_{i=1}^3s_{i}^{2}Y_{2}^{M}(\mathbf{\hat{s}}_{i})
\end{equation}
is the leading order term in the 111-expansion of the $\Psi$ we study in this paper.
$Y_l^m(\hat{\vect s})$ is the spherical harmonic defined in Ref.~\cite{macrobert1947}.
The superscript in $T^{(2)}$ means that $T^{(2)}$ is quadratic in $B$ for any fixed shape and orientation of the triangle
formed by the three bosons. The coefficient on the right hand side of \Eq{T2} is chosen such that
$T^{(2)}=\sum_{i=1}^3(s_{ix}+\I s_{iy})^2$ if $M=2$, where $s_{ix}$ and $s_{iy}$ are two of the three Cartesian components of $\vect s_i$.
We further require that the three bosons have well-defined orbital angular momentum quantum number $L=2$
and magnetic quantum number $M$, so that we can \emph{not} add a term proportional to the three-body wave function studied in Ref.~\cite{tan2008three} into the definition of $\Psi$ studied in this paper.
Then if the two-body potential $V(s)$ does not support any two-body bound state, $\Psi$ is uniquely defined.
But if $V(s)$ does support at least one bound state, then we need to have an additional condition to fix the definition of $\Psi$:
we specify that in the 21-expansion of $\Psi$ there may be terms describing the bound pair flying \emph{away} from the remaining free boson, but that there is \emph{no} term describing the bound pair flying \emph{toward} the free boson.

In order to prepare for the 21-expansion (which is intimately related to the 111-expansion), we will define the two-body special functions first. Similar two-body special functions were defined in Ref.~\cite{tan2008three}.

\subsection{Two-body special functions}

%In order to characterize the low-energy behaviour of ultracold atoms, we only need to concern several two-body parameters. 
We define the two-body special functions $\phi^{(l,m)}(\mathbf{s})$, $f^{(l,m)}(\mathbf{s})$, $g^{(l,m)}(\mathbf{s})$, $h^{(l,m)}(\mathbf{s})$, ..., which describe the collision of two particles with angular momentum quantum number $l$ and magnetic quantum number $m$. They are defined to satisfy
\begin{eqnarray}
    &&\widetilde{H} \phi^{(l,m)}(\mathbf{s})=0, \\
    &&\widetilde{H} f^{(l,m)}(\mathbf{s})=\phi^{(l,m)}(\mathbf{s})\label{eq:fs}, \\
    &&\widetilde{H} g^{(l,m)}(\mathbf{s})=f^{(l,m)}(\mathbf{s}), \\
    &&\widetilde{H} h^{(l,m)}(\mathbf{s})=g^{(l,m)}(\mathbf{s}).
\end{eqnarray}
$\hbar^2\widetilde{H}/M_B$ is the Hamiltonian for the relative motion of two bosons:
\begin{equation}
    \widetilde{H}=-\nabla_{\mathbf{s}}^2+\frac{M_B}{\hbar^2} V(s).
\end{equation}

As we stated before, the two-body potential vanishes at $s>r_e$. So $\phi^{(l,m)}(\vect s)$ is equal to
 $Y_l^m(\hat{\vect s})$ times a linear combination of $s^l$ and $s^{-l-1}$ at $s>r_e$.
We choose the overall amplitude of $\phi^{(l,m)}(\vect s)$ such that it takes the following form at $s>r_e$:
\begin{equation}
    \phi^{(l,m)}(\mathbf{s})=\left[\frac{s^l}{(2l+1)!!}-\frac{(2l-1)!!a_l}{s^{l+1}}\right]\sqrt{\frac{4\pi}{2l+1}}Y_{l}^{m}(\hat{\mathbf{s}}),\label{eq:phi}
\end{equation}
where $a_l$ is the scattering length or hypervolume of the $l$th partial wave and the dimension of $a_l$ is length to the $(2l+1)$st power.

There is some ``gauge freedom" in the definition of $f^{(l,m)}(\mathbf{s})$ since both $f^{(l,m)}(\mathbf{s})$ and $f^{(l,m)}(\mathbf{s})+c\phi^{(l,m)}(\mathbf{s})$ satisfy \Eq{eq:fs}, where $c$ is an arbitrary constant. 
To fix this ``gauge freedom", we require that in the expansion of $f^{(l,m)}(\mathbf{s})$ at $s>r_e$ we do \emph{not}
have a term proportional to $1/s^{l+1}$ [we have liberty to impose this condition on $f^{(l,m)}(\vect s)$ as long as $a_l\ne0$].
With this ``gauge condition", $f^{(l,m)}(\mathbf{s})$ takes the following form at $s>r_e$:
\begin{equation}
\begin{split}
        f^{(l,m)}(\mathbf{s})=&\biggl[-\frac{s^{l+2}}{2(2l+3)!!}-\frac{(2l-3)!!a_l}{2s^{l-1}}\\&-\frac{a_l r_l s^l}{2(2l+1)!!}\biggr]\sqrt{\frac{4\pi}{2l+1}}Y_{l}^{m}(\hat{\mathbf{s}}), \label{eq:f1}
\end{split}
\end{equation}
where $r_l$ is the $l$th partial wave effective range, whose dimension is length to the $(1-2l)$th power.

Similarly, we can derive the analytical expressions of the two-body special functions $g^{(l,m)}(\mathbf{s})$ and $h^{(l,m)}(\mathbf{s})$ at $s>r_e$:
\begin{equation}
    \begin{split}
        g^{(l,m)}(\mathbf{s})=&\biggl[\frac{s^{l+4}}{8(2l+5)!!}-\frac{(2l-5)!!a_l}{8s^{l-3}}+\frac{a_{l}r_{l}s^{l+2}}{4(2l+3)!!}\\&
        -\frac{a_l r'_{l}s^{l}}{24(2l+1)!!}\biggr]\sqrt{\frac{4\pi}{2l+1}}Y_{l}^{m}(\hat{\mathbf{s}}),\label{eq:g}
    \end{split}
\end{equation}
\begin{align}
        h^{(l,m)}(\mathbf{s})&=\biggl[-\frac{s^{l+6}}{48(2l+7)!!}-\frac{a_l (2l-7)!!}{48 s^{l-5}}-\frac{a_l r_l s^{4+l}}{16(2l+5)!!}
        \nonumber\\
        &\quad+\frac{a_l r'_{l}s^{l+2}}{48(2l+3)!!}-\frac{a_l r''_{l}s^l}{720(2l+1)!!}\biggr]\sqrt{\frac{4\pi}{2l+1}}Y_{l}^{m}(\hat{\mathbf{s}}). \label{eq:h}
\end{align}
$r'_l$ and $r''_l$ in Eqs.~\eqref{eq:g} and \eqref{eq:h} are called the shape parameters.
    
%We can derive the 21-expansion in terms of two-body special functions. 
Because of Bose statistics, $l$ must be even. We use the symbols $s$, $d$, and $g$ to represent $l=0$, $2$, and $4$, respectively.

The parameters $a_l$, $r_l$, $r'_l$, and $r''_l$ appear in the effective range expansion \cite{hammer2010causality}:
\begin{equation}\label{effectiverangeexpansion}
    k^{2l+1} \cot \delta_l = -\frac{1}{a_l}+\frac{1}{2} r_l k^2 +\frac{1}{4!} r'_{l} k^4 +\frac{1}{6!}r''_{l}k^{6}+O(k^8),
\end{equation}
where $\delta_l$ is the scattering phase shift. To derive \Eq{effectiverangeexpansion}, we consider two bosons colliding with energy $\hbar^2k^2/M_B$, orbital angular momentum quantum number $l$, and magnetic quantum number $m$ in the center-of-mass frame, and their wave function $\Phi^{(l,m)}(\vect s)$ satisfies the Schrödinger equation
\begin{equation}
    \widetilde{H} \Phi^{(l,m)}(\mathbf{s})=k^{2}\Phi^{(l,m)}(\mathbf{s}).
\end{equation}
Treating the energy as a small parameter, and choosing a suitable overall amplitude for $\Phi^{(l,m)}(\mathbf{s})$, we can expand $\Phi^{(l,m)}(\mathbf{s})$ in powers of $k^2$:
\begin{align}\label{Phismallenergy}
    \Phi^{(l,m)}(\mathbf{s})=\,&\phi^{(l,m)}(\mathbf{s})+k^2 f^{(l,m)}(\mathbf{s}) +k^4 g^{(l,m)}(\mathbf{s})\nonumber\\
    &+k^6 h^{(l,m)}(\mathbf{s}) +\cdots.
\end{align}
At $s>r_e$ we have
\begin{equation}\label{Phijy}
\Phi^{(l,m)}(\vect s)=A^{(l)}(k)\big[j_l(ks)\cot\delta_l-y_l(ks)\big],
\end{equation}
where $j_l$ and $y_l$ are the spherical Bessel functions, and $A^{(l)}(k)$ is a function of $l$ and $k$. Comparing \Eq{Phijy} with \Eq{Phismallenergy} and using Eqs.~\eqref{eq:phi}, \eqref{eq:f1}, \eqref{eq:g}, and \eqref{eq:h}, one can derive \Eq{effectiverangeexpansion}.

\subsection{111-expansion and 21-expansion}
%In this section, we will derive two asymptotic expansions of the three-body wave function. 
When the three pairwise distances $s_1$, $s_2$ and $s_3$ go to infinity simultaneously for any fixed ratio $s_1:s_2:s_3$, $\Psi$ can be expanded in powers of $1/B$. This expansion is called the 111-expansion. The 111-expansion can be written as
\begin{equation}
    \Psi = \sum_{n=-2}^{\infty} T^{(-n)}(\vect r_i, \vect r_j, \vect r_k),
\end{equation}
where $T^{(-n)}(\vect r_1, \vect r_2, \vect r_3)$ scales as $B^{-n}\ln^{p}B$ for some integer $p$ (which may depend on $n$) at large $B$. Since the pairwise distances $s_1$, $s_2$ and $s_3$ are large while the interactions have a finite range $r_e$, $T^{(-n)}$ satisfies
\begin{equation}
    -\frac{\hbar^2}{2 M_B}\left(\nabla_{1}^{2}+\nabla_{2}^{2}+\nabla_{3}^{2}\right) T^{(-n)}(\vect r_1, \vect r_2, \vect r_3)=0\label{eq:Tsc}
\end{equation}
if $s_1$, $s_2$, and $s_3$ are all nonzero.
If we extend the above equation to include the possibility of a vanishing $s_i$ (for $i=1$, $2$, or $3$),
we should add some terms that involve the delta function of $\vect s_i$ and the derivatives of such delta function.
If we extend the above equation to include the possibility of $\vect s_i=\vect R_i=0$ (i.e., all three bosons coinciding),
we should also add some terms that involve the derivatives of the six-dimensional delta function $\delta(\vect s_i)\delta(\vect R_i)$.

%If we want to discuss the behaviour of $T^{(n)}$ within the range of interaction, ~(\Eq{eq:Tsc}) should be equal to some delta functions and its derivatives (i.e. the summation of two-body and three-body sources).

When a boson, for instance the $i$th boson, is far away from the other two bosons and the distance of these two bosons (which are called the $j$th and the $k$th bosons) is fixed, $\Psi$ can be expanded in powers of $1/R_i$. This expansion is called the 21-expansion. 
The 21-expansion can be written as
\begin{equation}\label{Psi S}
    \Psi=\sum_{n=n_\text{min}}^{\infty}S^{(-n)}(\mathbf{R},\mathbf{s}),
\end{equation}
where $S^{(-n)}(\mathbf{R},\mathbf{s})$ scales as $R^{-n}$ (with the possibility of an extra factor which is a polynomial of $\ln R$),
and $n_\text{min}$ is an integer to be determined. 
$S^{(-n_\text{min})}$ is the leading-order nonzero term in the 21-expansion.
Without loss of generality, we omit the index $i$ in $\vect R_i$ and $\vect s_i$ in the 21-expansion.
When one boson is far away from the other two, \Eq{eq:sc} can be simplified as
\begin{equation}
    \left[-\nabla_{\mathbf{s}}^{2}-\frac{3}{4}\nabla_{\mathbf{R}}^{2}+\frac{M_B}{\hbar^{2}}V(s)\right]\Psi=0, \label{eq:S}
\end{equation}
which is valid if $R$ is large enough.
Substituting \Eq{Psi S} into \Eq{eq:S} and grouping the terms in powers of $1/R$, we derive the equations that $S^{(-n)}(\mathbf{R},\mathbf{s})$ satisfies:
\begin{eqnarray}
    &&\widetilde{H}S^{(-n_\text{min})}=0,\label{eq:S20} \\ 
    &&\widetilde{H}S^{(-n_\text{min}-1)}=0,\label{eq:S10} \\
    &&\widetilde{H}S^{(-n)}=\frac{3}{4}\nabla_{\mathbf{R}}^{2}S^{(-n+2)},\;n\geq n_\text{min}+2.
\end{eqnarray}
Considering the definition of the two-body special functions, we see that
$S^{(-n)}$ can be expressed as a linear combination of the two-body special functions for any value of $\vect R$, for each value of $n$.

In order to connect the 111-expansion and 21-expansion, we can expand $T^{(-n)}$ at $s\ll R$:
\begin{equation}
    T^{(-n)}=\sum_{i} t^{(i,-n-i)} ,\label{eq:Tt}
\end{equation}
where $t^{(i,-n-i)}$ scales as $R^{i}s^{-n-i}$ (with the possibility of an extra factor which is a polynomial of $\ln R$). Similarly, we can expand $S^{(-n)}$ at $r_e\ll s\ll R$ as
\begin{equation}
    S^{(-n)}=\sum_{j} t^{(-n,j)} .\label{eq:St}
\end{equation}
Note that the $t^{(i,j)}$ in the above two expansions should be the \emph{same} function for any given values of $i$ and $j$.

%Since three-body wave function $\Psi(\vect r_1,\vect r_2,\vect r_3)$ can be expanded at either $B\gg 1$ or $R\gg s$ limit, $t^{(m,n)}$ terms are identical in equation ~(\Eq{eq:Tt}) and ~(\Eq{eq:St}). 
$\Psi(\vect r_1,\vect r_2,\vect r_3)$ can be expressed as a double expansion at $r_e\ll s\ll R$:
\begin{equation}
    \Psi = \sum_{i,j} t^{(i,j)} .
\end{equation}

To derive the 111-expansion and 21-expansion, we start from the leading order term of the 111-expansion, \Eq{T2}.
%. In order to obtain a desired leading order term, we have the following requirements:
%\begin{itemize}
%    \item [a)]
%    It satisfies the three-body Schrödinger equation \Eq{eq:Tsc}.
%    \item [b)]
%    Since it is a bosonic system, the leading order term is invariant if the indices of the particle is changed.
%    \item [c)]
%    The system has translation symmetry(i.e. the leading order term is a function of $\vect s_i$ and $\vect r_i$).
%    \item[d)]
%    The leading order term should describe orbital angular momentum $L=2$.
%\end{itemize}

Since $T^{(-n)}=0$ at $n<-2$ for the wave function $\Psi$ that we study in this paper, we have
\begin{equation}\label{ti+j>2}
t^{(i,j)}=0,~~\text{if }i+j>2.
\end{equation}

Expanding $T^{(2)}$ at $s\ll R$, we get
\begin{eqnarray}
    &&t^{(2,0)}=8\sqrt{\frac{2\pi}{15}}R^{2}Y_{2}^{M}(\hat{\vect{R}}),\label{t20} \\
    &&t^{(0,2)}=6\sqrt{\frac{2\pi}{15}}s^{2}Y_{2}^{M}(\hat{\vect{s}}),\label{t02}\\
    &&t^{(i,2-i)}=0,~\text{if }i\ne2,0.
\end{eqnarray}

According to \Eq{eq:S20}, $S^{(-n_\text{min})}$ is a linear combination of $\phi^{(l,m)}(\vect s)$:
\begin{equation}
S^{(-n_\text{min})}=\sum_{l}\sum_{m=-l}^l c^{-n_\text{min}}_{l,m}\phi^{(l,m)}(\vect s),
\end{equation}
where $c^{-n_\text{min}}_{l,m}$ depends on $\vect R$ and scales like $R^{-n_\text{min}}$. Expanding $S^{(-n_\text{min})}$
at $r_e\ll s\ll R$, we find that $t^{(-n_\text{min},j)}\ne0$ for some $j\ge0$. Further considering \Eq{ti+j>2} and \Eq{t20},
we find that
\begin{equation}
n_\text{min}=-2,
\end{equation}
namely the leading-order nonzero term in the 21-expansion is $S^{(2)}$.

According to \Eq{eq:St} and \Eq{ti+j>2}, we have
\begin{equation}\label{S2 in terms of t}
    S^{(2)}=t^{(2,0)}+t^{(2,-1)}+t^{(2,-2)}+\cdots
\end{equation}
if $r_e\ll s\ll R$.
According to \Eq{eq:S20}, $S^{(2)}$ is a linear combination of $\phi^{(l,m)}(\mathbf{s})$, and the coefficients of such linear combination depend on $\vect R$.
Further considering the fact that in the expansion of $\phi^{(l,m)}(\mathbf{s})$ at $s>r_e$ there is a term proportional to $s^l$,
this linear combination can only involve $\phi^{(0,0)}(\vect s)$, or else the expansion of $S^{(2)}$ at $r_e\ll s\ll R$ would contain a term scaling as $R^2s^l$ with $l>0$ and this would violate \Eq{S2 in terms of t}. So we have
\begin{equation}\label{S2formula}
    S^{(2)}=c^{2}_{0,0}(\vect R)\phi^{(0,0)}\left(\mathbf{s}\right) ,
\end{equation}
where $c^{2}_{0,0}(\vect R)$ depends on $\vect R$. Expanding the right hand side of \Eq{S2formula} at $r_e\ll s\ll R$, and comparing
the resultant leading order term with \Eq{t20}, we get
\begin{equation}
c^{2}_{0,0}(\vect R)=8\sqrt{\frac{2\pi}{15}}R^{2}Y_{2}^{M}(\hat{\vect R}).
\end{equation}
Expanding $S^{(2)}$ at $r_e \ll s\ll R$, we get
\begin{align}
t^{(2,-1)}&=-8a\sqrt{\frac{2\pi}{15}}\frac{R^{2}}{s}Y_{2}^{M}(\hat{\vect R}),\\
t^{(2,j)}&=0,\text{ if }j\le-2,
\end{align}
where $a\equiv a_0\equiv a_s$ is the $s$-wave scattering length between two bosons.

Now we are able to calculate $T^{(1)}$. At $s\ll R$, we have
\begin{equation}
    T^{(1)}=t^{(2,-1)}+O(R^1s^0).
\end{equation}
$t^{(2,-1)}$ indicates the behavior of $T^{(-1)}$ when $s$ is small. Therefore $T^{(1)}$ satisfies
the following equation:
\begin{equation}
     \left(-\nabla_{\mathbf{s}_1}^{2}-\frac{3}{4}\nabla_{\mathbf{R}_1}^{2}\right)T^{(1)}=-\sum_{i=1}^3 32\sqrt{\frac{2\pi}{15}}\pi a R_i^{2} Y_{2}^{M}(\hat{\vect R}_i) \delta \left(\mathbf{s}_i\right) .\label{eq:T1}
\end{equation}
Solving \Eq{eq:T1} by using the Green function, requiring that $T^{(1)}$ scales like $B^1$, and considering the fact that there is \emph{no} nonzero harmonic polynomial scaling like $B^1$ and consistent with the complete symmetry of the wave function under interchanges of the bosons, we get a unique result for $T^{(1)}$:
\begin{equation}
T^{(1)}=\sum_{i=1}^{3}BP_{2,0}^{(1)}\gamma_{2,0}^{(1)}(\theta_{i})Y_{2}^{M}(\mathbf{\hat{R}}_{i})Y_{0}^{0}(\mathbf{\hat{s}}_{i}),
\end{equation}
where $P_{2,0}^{(1)}=-8\pi a\sqrt{\frac{2}{5}}$, $\gamma_{2,0}^{(1)}(\theta)=\tan\theta\sin\theta$. Expanding $T^{(1)}$ at $s \ll R$, we get
\begin{equation}
    t^{(1,0)}=-8\pi\sqrt{\frac{2}{15}}a R Y_{2}^{M}(\hat{\mathbf{R}})Y_{0}^{0}(\hat{\mathbf{s}})
\end{equation}
and explicit formulas for $t^{(i,1-i)}$ (which, for the sake of brevity, we do not show) for $i\le0$.

At $r_e\ll s\ll R$, we have
\begin{equation}\label{eq:S1re<<s}
S^{(1)}=t^{(1,1)}+t^{(1,0)}+O(R^1s^{-1}).
\end{equation}
Solving \Eq{eq:S10} (with $n_\text{min}=-2$) and \Eq{eq:S1re<<s}, we get
\begin{equation}
    S^{(1)}=c^{1}_{0,0}\phi^{(0,0)}\left(\mathbf{s}\right),
\end{equation}
where $c^{1}_{0,0}=-4\sqrt{\frac{2\pi}{15}}aRY_{2}^{M}(\hat{\mathbf{R}})$.
Expanding $S^{(1)}$ at $r_e\ll s\ll R$, we find 
\begin{equation}
t^{(1,-1)}=8\pi\sqrt{\frac{2}{15}}\frac{a^{2}R}{s}Y_{2}^{M}(\hat{\mathbf{R}})Y_{0}^{0}(\hat{\mathbf{s}}),
\end{equation}
and $t^{(1,j)}=0$ for $j\le-2$.

At $s\ll R$, we have
\begin{equation}
T^{(0)}=t^{(2,-2)}+t^{(1,-1)}+O(R^0s^0)=t^{(1,-1)}+O(R^0s^0).
\end{equation}
$t^{(1,-1)}$ indicates the behavior of $T^{(0)}$ when $s$ is small. Therefore $T^{(0)}$ satisfies
\begin{equation}
     \left(-\nabla_{\mathbf{s}_1}^{2}-\frac{3}{4}\nabla_{\mathbf{R}_1}^{2}\right)T^{(0)}=\sum_{i=1}^3 16\pi\sqrt{\frac{2\pi}{15}}a^{2}R_{i}Y_{2}^{M}(\hat{\vect R}_{i})\delta(\vect s_{i}).\label{eq:T0}
\end{equation}
Solving \Eq{eq:T0} by using the Green function, requiring that $T^{(0)}$ scales like $B^0$, we get
\begin{equation}
    T^{(0)}=T^{(0)}_0+\sum_{i=1}^{3}P_{2,0}^{(0)}\gamma^{(0)}_{2,0}(\theta_i)Y_{2}^{M}(\hat{\mathbf{R}}_{i})Y_{0}^{0}(\hat{\mathbf{s}}_{i}),
\end{equation}
where $T_{0}^{(0)}$ is a constant and satisfies the homogeneous differential equation $(-\nabla_{\mathbf{s}_1}^{2}-\frac{3}{4}\nabla_{\mathbf{R}_1}^{2})T^{(0)}_0=0$,
\begin{equation}
P^{(0)}_{2,0}=\frac{64}{15}\sqrt{\frac{2}{5}}a^{2},
\end{equation}
and we have introduced a mathematical function $\gamma_{l_{R},l_{s}}^{(n)}(\theta)$ (in which $l_R$ and $l_s$ are any nonnegative integers and $n$ is any integer):
\begin{widetext}
\begin{equation}
    \gamma_{l_{R},l_{s}}^{(n)}(\theta)\equiv\  _{2}F_{1}\left(\frac{1}{2}\left(l_{R}+l_{s}-n\right),\frac{1}{2}\left(4+l_{R}+l_{s}+n\right);l_{R}+\frac{3}{2};\sin^{2}\theta\right)\sin^{l_{R}}\theta\cos^{l_{s}}\theta,
\end{equation}    
where $_{2}F_{1}(a,b;c,z)$ is the hypergeometric function.
$\gamma_{l_{R},l_{s}}^{(n)}(\theta)$ satisfies the following differential equation:
\begin{equation}
\biggl[-\frac{d^{2}}{d\theta^{2}}-2\left(\cot\theta-\tan\theta\right)\frac{d}{d\theta}+\frac{l_{s}(l_{s}+1)}{\cos^{2}\theta}+\frac{l_{R}(l_{R}+1)}{\sin^{2}\theta}-n(4+n)\biggr]\gamma_{l_{R},l_{s}}^{(n)}(\theta)=0.
\end{equation}
\end{widetext}
When $\theta\to0$, $\gamma_{l_{R},l_{s}}^{(n)}(\theta)\propto\theta^{l_R}$, and this boundary condition is associated with
the analytical behavior of $T^{(n)}$ at $R_i\to0$ for any fixed value of $B$.
If $T^{(0)}_0\ne0$, $T^{(0)}_0$ is a term having orbital angular momentum quantum number $L=0$ in the three-body wave function. But we require that $\Psi$ is an eigenstate
of orbital angular momentum with $L=2$, so we must set $T^{(0)}_0=0$, and we get
\begin{equation}
    T^{(0)}=\sum_{i=1}^{3}P_{2,0}^{(0)}\gamma^{(0)}_{2,0}(\theta_i)Y_{2}^{M}(\hat{\mathbf{R}}_{i})Y_{0}^{0}(\hat{\mathbf{s}}_{i}).
\end{equation}
Then we expand $T^{(0)}$ at $s\ll R$ to determine the $t^{(i,j)}$ with $i+j=0$.
Then we compute $S^{(0)}$, $T^{(-1)}$, $S^{(-1)}$, and so on.
When we arrive at $T^{(-6)}$, we find that if $B>0$
\begin{widetext}
\begin{align}
    \left(-\nabla_{\vect s}^{2}-\frac{3}{4}\nabla_{\vect R}^{2}\right)T^{(-6)}=\sum_{i=1}^{3}\biggl[&\frac{5 \pi ^{3/2}}{8\sqrt{3} R_i^5}P_{2,0}^{(-6)}Y_{2}^{M}(\hat{\vect R}_i)\delta(\vect s_i) +\frac{\pi}{36R_i^{3}}\sqrt{\frac{5}{3}}P_{0,2}^{(-6)}Q_{2}^{M}(\nabla_{\vect s_i})\delta(\vect s_i)\nonumber\\&
    +\frac{49\pi ^{\frac{3}{2}}}{4R_i^3} \sqrt{15} P_{4,2}^{(-6)}\sum_{m}C^{2,M}_{4,m;2,M-m}Y_{4}^{m}(\hat{\vect R}_{i})Q_{2}^{M-m}(\nabla_{\vect s_i})\delta(\vect s_{i})\nonumber\\&
    +\frac{5}{48R_i^{3}} \sqrt{\frac{5}{3}} \pi ^{\frac{3}{2}}P_{2,2}^{(-6)}\sum_{m}C^{2,M}_{2,m;2,M-m}Y_{2}^{m}(\hat{\vect R}_{i})Q_{2}^{M-m}(\nabla_{s_i} )\delta(\vect s_i)\biggr],\label{eq:LaplaceT-6B>0}
\end{align}
where $C^{2,M}_{l,m;l',m'}$ is the Clebsch-Gordan coefficient, and
the coefficients $P^{(n)}_{l_{R},l_{s}}$ are listed in Appendix \ref{appendix:coefficients}.
Before using the Green function method to solve \Eq{eq:LaplaceT-6B>0} for $T^{(-6)}$, we need to extend \Eq{eq:LaplaceT-6B>0} to the origin (namely the point $\vect s_i=\vect R_i=0$) in the six-dimensional configuration space. At the origin there could be an effective source term proportional to $[Q_{2}^{M}(\nabla_{\vect s})+\frac{3}{4}Q_{2}^{M}(\nabla_{\vect R})]\delta(\vect s)\delta(\vect R)$ which is compatible with the requirement of Bose statistics and the requirement that $T^{(-6)}$ scales as $B^{-6}$ and so the source term scales like $B^{-8}$. Here
\begin{equation}
Q_{l}^{m}(\mathbf{r})\equiv\sqrt{\frac{4\pi}{2l +1}}r^{l}Y_{l}^{m}(\hat{\mathbf{r}})
\end{equation}
is a harmonic polynomial. The extension of \Eq{eq:LaplaceT-6B>0} to the entire configuration space is thus

\begin{align}
    \left(-\nabla_{\vect s}^{2}-\frac{3}{4}\nabla_{\vect R}^{2}\right)T^{(-6)}=&
    -\tilde{D}_\Lambda\Big[Q_{2}^{M}(\nabla_{\vect s})+\frac{3}{4}Q_{2}^{M}(\nabla_{\vect R})\Big]\delta(\vect s)\delta(\vect R)+\sum_{i=1}^{3}\biggl[\frac{5 \pi ^{3/2}}{24 R_i^5}Z_{\Lambda}(\vect R_{i})P_{2,0}^{(-6)}Y_{2}^{M}(\hat{\vect R}_i)\delta(\vect s_i)\nonumber \\&
    -\frac{\pi}{36R_i^{3}}\sqrt{\frac{5}{3}}Z_{\Lambda}(\vect R_i)P_{0,2}^{(-6)}Q_{2}^{M}(\nabla_{\vect s_i})\delta(\vect s_i)\nonumber\\&
    +\frac{49\pi ^{\frac{3}{2}}}{4R_i^3} \sqrt{15}Z(\vect R_i) P_{4,2}^{(-6)}\sum_{m}C^{2,M}_{4,m;2,M-m}Y_{4}^{m}(\hat{\vect R}_{i})Q_{2}^{M-m}(\nabla_{\vect s_i})\delta(\vect s_{i})\nonumber\\&
    +\frac{5}{48R_i^{3}} \sqrt{\frac{5}{3}} \pi ^{\frac{3}{2}}Z(\vect R_i)P_{2,2}^{(-6)}\sum_{m}C^{2,M}_{2,m;2,M-m}Y_{2}^{m}(\hat{\vect R}_{i})Q_{2}^{M-m}(\nabla_{s_i} )\delta(\vect s_i)\biggr],\label{eq:LaplaceT-6}
\end{align}
where $\tilde{D}_\Lambda$ is a new parameter that depends on the details of the two-body and three-body potentials, 
$\Lambda$ is a positive length scale whose value is arbitrary,
and $Z_{\Lambda}(\vect R)\frac{Y_l^m(\hat{\vect R})}{R^n}$ (where $n\ge3$ and $n-3-l$ is a nonnegative even integer)
and $Z(\vect R)\frac{Y_l^m(\hat{\vect R})}{R^n}$ (where $n\ge3$ and $n-3-l$ is \emph{not} a nonnegative even integer) are generalized functions (similar to those defined in Ref.~\cite{tan2008three}) defined as follows:
\begin{subequations}    
\begin{align}
    &Z_{\Lambda}(\vect R)\frac{Y_{l}^{m}(\hat{\vect R})}{R^{n}}=\frac{Y_{l}^{m}(\hat{\vect R})}{R^n},\quad\text{if}\; \vect R \neq 0,\\&
    \int_{\text{all}\;\vect R} d^{3}\vect R Z_{\Lambda}(\vect R)\frac{Y_{l}^{m}(\hat{\vect R})}{R^{n}}p_{l'}(\vect R)=0,\quad  \text{if}\; l'< n-3,\\&
    \int_{R< \Lambda}d^{3}\vect RZ_{\Lambda}(\vect R) \frac{Y_{l}^{m}(\hat{\vect R})}{R^{n}}p_{l'}(\vect R)=0,\quad \text{if}\;l'=n-3,\\&
    \int_{R<r_0}d^{3}\vect R Z_{\Lambda}(\vect R)\frac{Y_{l}^{m}(\hat{\vect R})}{R^{n}}p_{l'}(\vect R)=\int_{R< r_0}d^{3}\vect R \frac{Y_{l}^{m}(\hat{\vect R})}{R^{n}}p_{l'}(\vect R),\quad \text{if}\;l'> n-3,\\&
    Z(\vect R)\frac{Y_{l}^{m}(\hat{\vect R})}{R^{n}}=\frac{Y_{l}^{m}(\hat{\vect R})}{R^{n}},\quad\text{if}\; \vect R \neq 0,\\&
    \int_{\text{all}\;\vect R} d^{3}\vect R Z(\vect R)\frac{Y_{l}^{m}(\hat{\vect R})}{R^{n}}p_{l'}(\vect R)=0,\quad  \text{if}\; l'< n-3,\\&
    \int_{R< r_{0}}d^{3}\vect R Z(\vect R)\frac{Y_{l}^{m}(\hat{\vect R})}{R^{n}}p_{l'}(\vect R)=0,\quad \text{if}\;l'= n-3,\\&
    \int_{R< r_{0}}d^{3}\vect R Z(\vect R)\frac{Y_{l}^{m}(\hat{\vect R})}{R^{n}}p_{l'}(\vect R)=\int_{R< r_{0}}d^{3}\vect R \frac{Y_{l}^{m}(\hat{\vect R})}{R^{n}}p_{l'}(\vect R),\quad \text{if}\;l'> n-3,
\end{align}
\end{subequations}
where $p_{l'}(\vect R)$ is any homogeneous degree-$l'$ polynomial of $\vect R$, and $r_0$ is an arbitrary positive value.
If $P^{(-6)}_{2,0}$ or $P^{(-6)}_{0,2}$ is nonzero, then in general the value of $\tilde{D}_\Lambda$ should depend on $\Lambda$
in such a way that the right hand side of \Eq{eq:LaplaceT-6} does \emph{not} depend on the arbitrary length scale $\Lambda$:
\begin{equation}
\tilde{D}_\Lambda-\tilde{D}_{\Lambda'}=-\frac{\sqrt{15}\pi^{2}}{18}(P^{(-6)}_{2,0}+P^{(-6)}_{0,2})\ln\frac{\Lambda}{\Lambda'},
\end{equation}
where
\begin{equation}
P^{(-6)}_{2,0}+P^{(-6)}_{0,2}=J_1a^8+J_2a^7r_s+J_3a^6r_s^2+J_4a^5r_s'+J_5a^3a_d,
\end{equation}
and $J_1$, $J_2$, $J_3$, $J_4$, and $J_5$ are numerical coefficients whose analytical expressions are listed in Appendix~\ref{appendix:coefficients}.
Solving \Eq{eq:LaplaceT-6} using the Green function method, we find
\begin{align}
    T^{(-6)}=\sum_{i=1}^{3}\biggl\{& -4\sqrt{\frac{2\pi}{15}}s_{i}^{2}Y_{2}^{M}(\mathbf{\hat{s}}_{i})\frac{1}{ B^{8}}\Big[D_{\Lambda}-\frac{3}{32\pi}\sqrt{\frac{15}{2}}(P_{2,0}^{(-6)}+P_{0,2}^{(-6)})\ln \frac{B}{\Lambda}\Big]\nonumber\\&
    +B^{-6}P^{(-6)}_{2,2}\gamma_{2,2}^{(-6)}(\theta_{i})\sum_{m}C^{2,M}_{2,m;2,M-m}Y_{2}^{m}(\mathbf{\hat{R}}_{i})Y_{2}^{M-m}(\mathbf{\hat{s}}_{i})\nonumber\\&
    +B^{-6}P^{(-6)}_{4,2}\gamma_{4,2}^{(-6)}(\theta_{i})\sum_{m}C^{2,M}_{4,m;2,M-m}Y_{4}^{m}(\mathbf{\hat{R}}_{i})Y_{2}^{M-m}(\mathbf{\hat{s}}_{i})\nonumber\\&
    +B^{-6}P_{2,0}^{(-6)}\left[{\dot\gamma}_{2,0}^{(-6)}(\theta)\right]Y_{2}^{M}(\mathbf{\hat{R}}_{i})Y_{0}^{0}(\mathbf{s}_{i})+B^{-6}P_{0,2}^{(-6)}\left[{\dot\gamma}_{0,2}^{(-6)}(\theta)\right]Y_{0}^{0}(\mathbf{\hat{R}}_{i})Y_{2}^{M}(\mathbf{s}_{i})\biggr\},
\end{align}
where $D_\Lambda\equiv\frac{27\sqrt{2}}{32\pi^{3}}\tilde{D}_\Lambda$, and
\begin{subequations}
\begin{equation}
{\dot\gamma}_{2,0}^{(-6)}(\theta)\equiv \frac{\partial\gamma_{2,0}^{(n)}(\theta)}{\partial n}\Big|_{n=-6}=
-\frac{\theta}{32}  (5 \cos 2 \theta -4 \cos 4 \theta +\cos 6 \theta ) \csc ^3\theta  \sec \theta +\frac{1}{240} \left(114-74 \cos 2 \theta +15 \csc ^2\theta \right),
\end{equation}
\begin{equation}
\dot\gamma^{(-6)}_{0,2}(\theta)\equiv\frac{\partial\gamma^{(n)}_{0,2}(\theta)}{\partial n}\Big|_{n=-6}
=-\frac{\theta}{32}  (5 \cos 2 \theta +4 \cos 4 \theta +\cos 6 \theta ) \csc \theta  \sec ^3\theta +\frac{1}{96} (17+12 \cos 2 \theta +\cos 4 \theta) \sec ^2 \theta.
\end{equation}
\end{subequations}
Then we expand $T^{(-6)}$ at $s\ll R$ to determine $t^{(-6,0)}$. Then we compute $S^{(-6)}$ (which is the order $R^{-6}$ term in the 21-expansion of $\Psi$). In order to express the 111-expansion with a simpler form, we imply the bipolar harmonics $Y_{l_{R},l_{s}}^{L,M}(\hat{\mathbf{R}},\hat{\mathbf{s}})$, which is defined as
\begin{equation}
    Y_{l_{R},l_{s}}^{L,M}(\hat{\mathbf{R}},\hat{\mathbf{s}})=\sum_{m}C^{L,M}_{l_{R},m;l_{s},M-m}Y_{l_{R}}^{m}(\hat{\mathbf{R}})Y_{l_{s}}^{M-m}(\hat{\mathbf{s}})
\end{equation}

Our final result for the 111-expansion of $\Psi$ is

\begin{align}
    \Psi=&\sum_{i=1}^{3}\biggl\{ 4\sqrt{\frac{2\pi}{15}}s_{i}^{2}Y_{2}^{M}(\mathbf{\hat{s}}_{i})\Big\{1-\frac{1}{ B^{8}}\Big[D_{\Lambda}-\frac{3}{32\pi}\sqrt{\frac{15}{2}}(P_{2,0}^{(-6)}+P_{0,2}^{(-6)})\ln \frac{B}{\Lambda}\Big]\Big\}+B^{1}P_{2,0}^{(1)}\gamma_{2,0}^{(1)}(\theta_{i})Y_{2,0}^{2,M}(\hat{\mathbf{R}}_{i},\hat{\mathbf{s}}_{i})\nonumber\\&
    +B^{0}P_{2,0}^{(0)}\gamma_{2,0}^{(0)}(\theta_{i})Y_{2,0}^{2,M}(\hat{\mathbf{R}}_{i},\hat{\mathbf{s}}_{i})+B^{-1}P_{2,0}^{(-1)}\gamma_{2,0}^{(-1)}(\theta_{i})Y_{2,0}^{2,M}(\hat{\mathbf{R}}_{i},\hat{\mathbf{s}}_{i})+B^{-2}P_{2,0}^{(-2)}\gamma_{2,0}^{(-2)}(\theta_{i})Y_{2,0}^{2,M}(\hat{\mathbf{R}}_{i},\hat{\mathbf{s}}_{i})\nonumber\\&
    +B^{-3}P_{2,0}^{(-3)}\gamma_{2,0}^{(-3)}(\theta_{i})Y_{2,0}^{2,M}(\hat{\mathbf{R}}_{i},\hat{\mathbf{s}}_{i})
    +B^{-3}P_{0,2}^{(-3)}\gamma_{0,2}^{(-3)}(\theta_{i})Y_{0,2}^{2,M}(\hat{\mathbf{R}}_{i},\hat{\mathbf{s}}_{i})+B^{-4}P_{2,0}^{(-4)}\gamma_{2,0}^{(-4)}(\theta_{i})Y_{2,0}^{2,M}(\hat{\mathbf{R}}_{i},\hat{\mathbf{s}}_{i})\nonumber\\&
    +B^{-4}P_{0,2}^{(-4)}\gamma_{0,2}^{(-4)}(\theta_{i})Y_{0,2}^{2,M}(\hat{\mathbf{R}}_{i},\hat{\mathbf{s}}_{i})+B^{-4}P_{2,2}^{(-4)}\gamma_{2,2}^{(-4)}(\theta_{i})Y_{2,2}^{2,M}(\hat{\mathbf{R}}_{i},\hat{\mathbf{s}}_{i})
    +B^{-4}P_{4,2}^{(-4)}\gamma_{4,2}^{(-4)}(\theta_{i})Y_{4,2}^{2,M}(\hat{\mathbf{R}}_{i},\hat{\mathbf{s}}_{i})\nonumber\\&
    +B^{-5}P_{2,0}^{(-5)}\gamma_{2,0}^{(-5)}(\theta_{i})Y_{2,0}^{2,M}(\hat{\mathbf{R}}_{i},\hat{\mathbf{s}}_{i})
    +B^{-5}P_{0,2}^{(-5)}\gamma_{0,2}^{(-5)}(\theta_{i})Y_{0,2}^{2,M}(\hat{\mathbf{R}}_{i},\hat{\mathbf{s}}_{i})+B^{-5}P_{2,2}^{(-5)}\gamma_{2,2}^{(-5)}(\theta_{i})Y_{2,2}^{2,M}(\hat{\mathbf{R}}_{i},\hat{\mathbf{s}}_{i})\nonumber\\&
    +B^{-5}P_{4,2}^{(-5)}\gamma_{4,2}^{(-5)}(\theta_{i})Y_{4,2}^{2,M}(\hat{\mathbf{R}}_{i},\hat{\mathbf{s}}_{i})
    +B^{-6}P_{2,2}^{(-6)}\gamma_{2,2}^{(-6)}(\theta_{i})Y_{2,2}^{2,M}(\hat{\mathbf{R}}_{i},\hat{\mathbf{s}}_{i})
    +B^{-6}P_{4,2}^{(-6)}\gamma_{4,2}^{(-6)}(\theta_{i})Y_{4,2}^{2,M}(\hat{\mathbf{R}}_{i},\hat{\mathbf{s}}_{i})\nonumber\\&
    +B^{-6}P_{2,0}^{(-6)}{\dot\gamma}_{2,0}^{(-6)}(\theta_i)Y_{2,0}^{2,M}(\hat{\mathbf{R}}_{i},\hat{\mathbf{s}}_{i})
    +B^{-6}P_{0,2}^{(-6)}{\dot\gamma}_{0,2}^{(-6)}(\theta_i)Y_{0,2}^{2,M}(\hat{\mathbf{R}}_{i},\hat{\mathbf{s}}_{i})\biggr\}
    +O(B^{-7}\ln^{n}B),\label{eq:111}
\end{align}
where $n$ is a nonnegative integer, and $D\equiv D_\Lambda$. %and $\gamma_{B\; l_{R},l_{s}}^{(n)}(\theta)=(B/\Lambda)^{n}\gamma_{l_R,l_s}^{(n)}(\theta)$
Explicit formulas for the functions $\gamma^{(n)}_{l_{R},l_{s}}(\theta)$ in the above expansion are listed in Appendix \ref{ref:function}. %${\dot\gamma}_{2,0}^{(-6)}(\theta)$ and ${\dot\gamma}_{0,2}^{(-6)}(\theta)$ are defined as follows:
%\begin{equation}
%{\dot\gamma}_{2,0}^{(-6)}(\theta)\equiv \frac{\partial\gamma_{2,0}^{(n)}(\theta)}{\partial n}\Big|_{n=-6}=
%-\frac{\theta}{32}  (5 \cos 2 \theta -4 \cos 4 \theta +\cos 6 \theta ) \csc ^3\theta  \sec \theta +\frac{1}{240} \left(114-74 \cos 2 \theta +15 \csc ^2\theta \right),
%\end{equation}
%\begin{equation}
%\dot\gamma^{(-6)}_{0,2}(\theta)\equiv\frac{\partial\gamma^{(n)}_{0,2}(\theta)}{\partial n}\Big|_{n=-6}
%=-\frac{\theta}{32}  (5 \cos 2 \theta +4 \cos 4 \theta +\cos 6 \theta ) \csc \theta  \sec ^3\theta +\frac{1}{96} (17+12 \cos 2 \theta +\cos 4 \theta) \sec ^2 \theta.
%\end{equation}
%The   term $T_{\text{3-body}}$ in \Eq{eq:111} proportional to $D$ is the solution of the following equation:
%\begin{equation}
%    \left(-\nabla_{\vect s}^{2}-\frac{3}{4}\nabla_{\vect R}^{2}\right)T_{\text{3-body}}=-\frac{32 D \pi^{3}}{27\sqrt{2}}\left[Q_{2}^{m}(\nabla_{\vect s})+\frac{3}{4}Q_{2}^{m}(\nabla_{\vect R})\right]\delta(\vect s)\delta(\vect R)
%\end{equation}
%where $Q_{l}^{m}(\mathbf{r})\equiv\sqrt{\frac{4\pi}{2l +1}}r^{l}Y_{l}^{m}(\hat{\mathbf{r}})$.
Our result for the 21-expansion of $\Psi$ (in the absence of two-body bound states) is
\begin{align}
    \Psi=&c_{0,0}^{2}(\vect R)\phi^{(0,0)}\left(\mathbf{s}\right)+c_{0,0}^{1}(\vect R)\phi^{(0,0)}\left(\mathbf{s}\right)+c_{2,M}^{0}(\vect R)\phi^{(2,M)}\left(\mathbf{s}\right)+\sum_{l=0,2}\sum_{m}c_{l,m}^{-1}(\vect R)\phi^{(l,m)}\left(\mathbf{s}\right)+\sum_{l=0,2}\sum_{m}c_{l,m}^{-2}(\vect R)\phi^{(l,m)}\left(\mathbf{s}\right) \nonumber\\&
    +\sum_{l=0,2,4}\sum_{m}c_{l,m}^{-3}(\vect R)\phi^{(l,m)}\left(\mathbf{s}\right)+\sum_{l=0,2,4}\sum_{m}c_{l,m}^{-4}(\vect R)\phi^{(l,m)}\left(\mathbf{s}\right)+\sum_{l=0,2,4,6}\sum_{m}c_{l,m}^{-5}(\vect R)\phi^{(l,m)}\left(\mathbf{s}\right)\nonumber\\&+\sum_{l=0,2,4,6}\sum_{m}c_{l,m}^{-6}(\vect R)\phi^{(l,m)}\left(\mathbf{s}\right)
    +\tilde{c}^{1}_{0,0}(\vect R)f^{(0,0)}\left(\mathbf{s}\right)+\tilde{c}_{0,0}^{0}(\vect R)f^{(0,0)}\left(\mathbf{s}\right)+\sum_{l=0,2}\sum_{m}\tilde{c}_{l,m}^{-1}(\vect R)f^{(l,m)}\left(\mathbf{s}\right)\nonumber\\&
    +\sum_{l=0,2}\sum_{m}\tilde{c}_{l,m}^{-2}(\vect R)f^{(l,m)}\left(\mathbf{s}\right)
    +\sum_{l=2,4}\sum_{m}\tilde{c}_{l,m}^{-3}(\vect R)f^{(l,m)}\left(\mathbf{s}\right)
    +\sum_{l=0,2,4}\sum_{m}\tilde{c}_{l,m}^{-4}(\vect R)f^{(l,m)}\left(\mathbf{s}\right)
    +\tilde{\tilde{c}}_{0,0}^{1}(\vect R)g^{(0,0)}\left(\mathbf{s}\right)\nonumber\\&
    +\tilde{\tilde{c}}_{0,0}^{0}(\vect R)g^{(0,0)}\left(\mathbf{s}\right)+\sum_{m}\tilde{\tilde{c}}_{2,m}^{-1}(\vect R)g^{(2,m)}\left(\mathbf{s}\right)+\sum_{l=0,2}\sum_{m}\tilde{\tilde{c}}_{l,m}^{-2}(\vect R)g^{(l,m)}\left(\mathbf{s}\right)
    +\tilde{\tilde{\tilde{c}}}_{0,0}^{0}(\vect R)h^{(0,0)}\left(\mathbf{s}\right)+O\big(R^{-7}\ln^{n_{1}}R\big),
    \label{eq:21expansionelastic}
\end{align}
where $n_{1}$ is a nonnegative integer. The functions $c_{l,m}^{i}(\vect R)$ scale as $R^i$, and their formulas are listed in Appendix \ref{appendix:coefficients}.
The functions $\tilde{c}_{l,m}^{i}(\vect R)$, $\tilde{\tilde{c}}_{l,m}^{i}(\vect R)$, and $\tilde{\tilde{\tilde{c}}}_{l,m}^{i}(\vect R)$ are defined as follows:
\end{widetext}
\begin{eqnarray}
    &&\tilde{c}_{l,m}^{i}(\vect R)=\frac{3}{4}\nabla^{2}_{\mathbf{R}}c_{l,m}^{i}(\vect R),\\
    &&\tilde{\tilde{c}}_{l,m}^{i}(\vect R)=\frac{3}{4}\nabla^{2}_{\mathbf{R}}\tilde{c}_{l,m}^{i}(\vect R),\\
    &&\tilde{\tilde{\tilde{c}}}_{l,m}^{i}(\vect R)=\frac{3}{4}\nabla^{2}_{\mathbf{R}}\tilde{\tilde{c}}_{l,m}^{i}(\vect R).
\end{eqnarray}
$\tilde{c}_{l,m}^{i}(\vect R)$ scales as $R^{i-2}$, $\tilde{\tilde{c}}_{l,m}^{i}(\vect R)$ scales as $R^{i-4}$,
and $\tilde{\tilde{\tilde{c}}}_{l,m}^{i}(\vect R)$ scales as $R^{i-6}$.

%$\overset{\cdot}{\gamma}_{2,0}^{(-6)}(\theta)$ and $\overset{\cdot}{\gamma}_{0,2}^{(-6)}(\theta)$ 

Equation~\eqref{eq:21expansionelastic} is valid if the two-body potential $V(s)$ does \emph{not} support any bound state.
If the two-body potential does support one or more bound states, the 21-expansion should be modified as
\begin{equation}
    \Psi = \Phi_\text{elastic} + \sum_{i=1}^{3}\sum_{l_1,l,m,\nu} c_{l_1,l,\nu}C^{2,M}_{l_1,M-m;l,m}\phi_{i,l_1,l,\nu}^{(m)}\left(\vect R_i,\vect s_i\right),
\end{equation}
where $\Phi_\text{elastic}$ is the right hand side of \Eq{eq:21expansionelastic},
%$C^{2,M}_{l_1,M-m;l,m}$ is the Clebsch-Gordan coefficient,
and
\begin{align}
    \phi_{i,l_1,l,\nu}^{(m)}\left(\vect R_i,\vect s_i\right)&=h^{(1)}_{l_1}\left(\frac{2 \kappa_{l,\nu}R_i}{\sqrt{3}}\right) u_{l,\nu}\left(s_i\right)Y_{l}^{m}\left(\hat{\vect s}_i\right)\nonumber\\
&\quad\times Y_{l_1}^{M-m}\left(\hat{\vect R}_i\right)
\end{align}
describes the process of a two-body bound state flying apart from the remaining free boson.
Here $h^{(1)}_{l_1}\left(z\right)$ is the Hankel function of the first kind. 
$u_{l,\nu}(s_i)Y_{l}^m(\hat{\vect s}_i)$ is the wave function of the two-body bound state with orbital angular momentum quantum number $l$, magnetic quantum quantum number $m$, vibrational quantum number $\nu$, and energy $-\hbar^2\kappa_{l,\nu}^2/m$, where $\kappa_{l,\nu}>0$ \cite{shinazhu2017}. 
%$\phi_{i,l,l_1,\nu}^{(m)}\left(\vect s_i,\vect r_i\right)$ is the bound state wave function which can be written as .
The radial function $u_{l,\nu}\left(s_i\right)$ is normalized such that
\begin{equation}
    \int u_{l,\nu}(s)u_{l',\nu'}^{*}(s)Y_{l}^{m}(\hat{\vect s})Y_{l'}^{m'}(\hat{\vect s})^{*}d^{3}\mathbf{s}
    =\frac{\delta_{l,l'}\delta_{m,m'}\delta_{\nu,\nu'}}{\kappa_{l,\nu}^{7}} ,
\end{equation}
where $\delta_{l,l'}$ is the Kronecker delta. From the conservation of probability \cite{petrov2019}, we derive
a relation between the imaginary part of $D$ and the coefficients $c_{l_1,l,\nu}$:
\begin{equation}
    \mathrm{Im}D=-\frac{81}{32\pi^{2}}\sum_{l_1,l,\nu}\frac{|c_{l_1,l,\nu}|^{2}}{\kappa_{l,\nu}^{8}}.
\end{equation}

\section{approximate formula of D for weak interactions \label{sec:3}}
For weak interactions, we can expand the wave function as a Born series:
\begin{equation}
    \Psi = \Psi_{0}+\hat{G}U\Psi_{0}+(\hat{G}U)^{2}\Psi_{0}+\cdots ,
\end{equation}
where $\hat{G}=-\frac{1}{\hat{H}_0}$ is the Green operator, $\hat{H}_0=-\frac{\hbar^2}{2M_B}(\nabla_1^2+\nabla_2^2+\nabla_3^2)$ is the three-body kinetic energy operator, and 
\begin{equation}
U=V_\text{3-body}\left(s_1,s_2,s_3\right)+\sum_{i=1}^{3}V\left(s_i\right)
\end{equation}
is the total interaction potential. $\Psi_{0}$ corresponds to the leading order term of the 111-expansion, namely the $T^{(2)}(\vect r_1,\vect r_2,\vect r_3)$ defined in \Eq{T2}:
\begin{equation}
    \Psi_{0}=4\sqrt{\frac{2\pi}{15}}\sum_{i=1}^3s_{i}^{2}Y_{2}^{M}(\mathbf{\hat{s}}_{i}).
\end{equation}
We then compute the first order term $\hat{G}U\Psi_0$ in the Born series. See Appendix~\ref{app:Born} for details.
When $s_1$, $s_2$, and $s_3$ are all large, we get
\begin{widetext}
\begin{equation}
\begin{split}
    \hat{G}U\Psi_{0}\approx&-4\sqrt{\frac{2\pi}{15}}\sum_{i=1}^{3}\left[2R_{i}^{2}Y_{2}^{M}(\hat{\vect R}_i)\left(\frac{\alpha_{1}}{s_i}\right)+\frac{3}{10}s_{i}^{2}Y_{2}^{M}(\hat{\vect s}_i)\left(\frac{\alpha_{5}}{s_{i}^{5}}\right)\right]
    -\frac{4\Delta}{ B^{8}}\sqrt{\frac{2\pi}{15}}\sum_{i=1}^3s_{i}^{2}Y_{2}^{M}(\mathbf{\hat{s}}_{i}),
\end{split}
\end{equation}
\begin{equation}
    \Delta\equiv\frac{\sqrt{3}M_{B}}{20 \pi \hbar^{2}}\int_{0}^\infty dR\int_{0}^\infty ds \int_{0}^{\pi}d\theta V(s,R,\theta) R^2 s^2 \sin \theta \left(16 R^4+18 R^2 s^2 \cos 2 \theta +6 R^2 s^2+9 s^4\right),\label{eq:Lambda}
\end{equation}
\end{widetext}
where $\theta$ is the angle between $\vect R$ and $\vect s$ and
\begin{eqnarray}
     &&\alpha_{n}=\frac{M_B}{\hbar^{2}}\int_{0}^{\infty}ds' s^{' n+1}V(s'), \\
    &&\alpha_{n}\left(s_{i}\right)=\frac{M_B}{\hbar^{2}}\int_{0}^{s_{i}}ds' s^{' n+1}V(s'), \\
    &&a_{n}\left(s_{i}\right)=\frac{M_B}{\hbar^{2}}\int_{s_{i}}^{\infty}ds' s^{' n+1}V(s'). 
\end{eqnarray}
By comparing the above results with the 111-expansion of $\Psi$, we obtain the expressions of the three-body scattering hypervolume $D$, scattering length $a$ and two-body $d$-wave scattering hypervolume $a_{d}$ to the first order in the interaction potential:
\begin{align}
    D&\approx\Delta,\label{DBorn}\\
    a&\approx\alpha_1 ,\\
    a_{d}&\approx \frac{2}{450}\alpha_{5}.
\end{align}
Note that $D$ depends on the three-body potential at leading order in the Born series expansion, according to \Eq{DBorn}.
Inspired by Ref.~\cite{shinawang2022}, we expect that $D$ depends also on the two-body potential $V(s)$ at second order in the Born series expansion.

\section{energy shift of three bosons with different momenta in a large periodic box \label{sec:4}}
In this section, we consider the energy shift of three bosons caused by the scattering hypervolumes in a large periodic cubic box with volume $\Omega$. If we neglect the effects of interactions,
the wave function of three free bosons with three \emph{different} momenta $\hbar \vect k_1$, $\hbar \vect k_2$, and $\hbar \vect k_3$ can be written as
\begin{equation}
    \Psi_\text{free}=\frac{1}{\sqrt{6}\Omega^{\frac{3}{2}}}\sum_{P}e^{\sum_{i=1}^{3}\I \mathbf{k}_{Pi}\cdot \vect r_i},
    \label{eq:periodicwf}
\end{equation}
where %$n_{i}$ is the number of particles which has the momentum $k_{i}$ and 
$P$ is any one of the six permutations of the integers $1,2,3$. %There are six terms on the right side of \Eq{eq:periodicwf}.

We define the Jacobi momenta $\hbar \mathbf{p}, \hbar \mathbf{q}, \hbar \mathbf{k_c}$:
\begin{subequations}\label{pq}
\begin{eqnarray}
    &&\vect k_1=\frac{1}{3}\mathbf{k_c}+\frac{1}{2}\mathbf{q}+\mathbf{p}, \\
    &&\vect k_2=\frac{1}{3}\mathbf{k_c}+\frac{1}{2}\mathbf{q}-\mathbf{p}, \\
    &&\vect k_3=\frac{1}{3}\mathbf{k_c}-\mathbf{q}.
\end{eqnarray}
\end{subequations}
Assuming that $k_1\sim k_2\sim k_3\sim1/\lambda$,
we expand $\Psi_\text{free}$ at $B\ll \lambda$:
%Noted that the correspond expansion is a second order homogeneous polynomial, which means it can be expressed as the linear combination of the second order harmonic polynomials and $B^2$.
\begin{equation}
    \Psi_\text{free}=\frac{e^{\I \vect k_{c}\cdot \vect r_{c}}}{\Omega^{\frac{1}{2}}}
    \big[\Phi^{(0)}_\text{free}-\frac{1}{6\sqrt{6}\Omega}(4p^{2}+3q^{2})B^2+\Phi^{(2)}_\text{free}+O(B^3)\big],
\end{equation}
where $\vect r_c=(\vect r_1+\vect r_2+\vect r_3)/3$ is the center-of-mass position vector,
\begin{equation}
\Phi^{(0)}_\text{free}=\frac{\sqrt{6}}{\Omega}
\end{equation}
is rotationally invariant (ie, belonging to orbital angular momentum quantum number $L=0$), and
\begin{align}
    \Phi^{(2)}_\text{free}=-\sum_{M=-2}^{2}\frac{1}{3\sqrt{6}\Omega}
    &\left[4Q_{2}^{-M}(\mathbf{p})+3Q_{2}^{-M}(\mathbf{q})\right]\nonumber\\
    &\times\Big[Q^{M}_{2}(\vect R_3)+\frac{3}{4}Q_{2}^{M}(\vect s_{3})\Big]
\end{align}
has orbital angular momentum quantum number $L=2$.
If we introduce the $L=0$ scattering hypervolume $D^{(0)}$ (defined in Ref.\cite{tan2008three}) and the $L=2$ scattering hypervolume $D$ (defined in this paper) adiabatically but still neglect the effects of the two-body interactions such as the $s$-wave scattering length $a$, the three-body wave function $\Psi_\text{free}$ is changed to
\begin{equation}
\Psi_\text{interacting}=\frac{e^{\I \vect k_{c}\cdot \vect r_{c}}}{\Omega^{\frac{1}{2}}}\Phi\label{eq:Psiinteracting},
\end{equation}
where the wave function for relative motion, $\Phi$, has a partial wave expansion at $r_e\ll B\ll\lambda$:
\begin{align}
    \Phi\approx &\Phi'+\frac{\sqrt{6}}{\Omega}\left(1-\frac{\sqrt{3}D^{(0)}}{8\pi^3 B^4}\right)-\frac{1}{6\sqrt{6}\Omega}(4p^{2}+3q^{2})B^2\nonumber\\&
    -\sum_{M=-2}^{2}\frac{1}{3\sqrt{6}\Omega}
    \left[4Q_{2}^{-M}(\mathbf{p})+3Q_{2}^{-M}(\mathbf{q})\right]\nonumber\\&\quad\quad\quad\quad\times
    \Big[Q^{M}_{2}(\vect R_3)+\frac{3}{4}Q_{2}^{M}(\vect s_{3})\Big]
    \left(1-\frac{D}{ B^{8}}\right),\label{eq:energywavefunction}
\end{align}
where $\Phi'$ is the sum of other partial waves ($L\ne0,2$) that are not considered explicitly. %The remaining four terms are similar since the scattering hypervolume is identical for different magnetic quantum number. These five terms are summed and denoted as $\Phi$.
The $\Phi$ in \Eq{eq:Psiinteracting} satisfies the free Schrödinger equation outside of the range of interaction:
\begin{equation}
    -\frac{\hbar^{2}}{M_B}\nabla^{2}_{\mathbf{\xi}}\Phi_{\vect p, \vect q}=E\Phi_{\vect p, \vect q},
\end{equation}
where $\mathbf{\xi}=(\mathbf{s_{3}},\frac{2}{\sqrt{3}}\mathbf{R_{3}})$ and $E$ is the energy of relative motion
[if there is no interaction, then $E=\hbar^2(k_1^2+k_2^2+k_3^2)/2M_B-\hbar^2k_c^2/6M_B$]. If we change the interactions adiabatically such that $D^{(0)}$ and $D$ are changed but the two-body parameters (including $a$, $r_s$, $r_s'$, and $a_d$) are not changed, the three-body energy eigenvalue will be changed. We denote such change of the energy as the energy shift. In order to calculate this energy shift, we can write the Schrödinger equation for relative motion for two different sets of interactions, whose corresponding wave functions for relative motion are denoted as $\Phi_1$ and $\Phi_2$, and whose energies for relative motion are denoted as $E_1$ and $E_2$:
\begin{eqnarray}
        &&-\frac{\hbar^{2}}{M_B}\nabla^{2}_{\mathbf{\xi}}\Phi_{1}=E_{1}\Phi_{1}, \label{eq:E1}\\ 
        &&-\frac{\hbar^{2}}{M_B}\nabla^{2}_{\mathbf{\xi}}\Phi_{2}=E_{2}\Phi_{2} ,\label{eq:E2}
\end{eqnarray}
where $\Phi_{1}$ contains three-body scattering hypervolumes $D^{(0)}_{1}$ and $D_{1}$, and $\Phi_{2}$ contains three-body scattering hypervolumes $D^{(0)}_{2}$ and $D_{2}$. From Eqs.~\eqref{eq:E1} and \eqref{eq:E2}, we derive
\begin{equation}
\begin{split}
    &(E_{1}-E_{2})\int_{\xi > \xi_{0}}d^{6}\xi\Phi_{1}\Phi_{2}^{*}=\\&
    -\frac{\hbar^{2}}{M_B}\int_{\xi > \xi_{0}} d^{6}\xi \nabla_{\mathbf{\xi}}\cdot \left(\Phi_{2}^{*}\nabla_{\xi}\Phi_{1}-\Phi_{1}\nabla_{\mathbf{\xi}}\Phi_{2}^{*}\right), \label{eq:energy}
\end{split}
\end{equation}
where $\xi_0$ is a length scale satisfying $r_e\ll\xi_{0}\ll\lambda$.
The left hand side of \Eq{eq:energy} can be calculated approximately with the assumption that $\xi_{0}$ is negligible compared with the size of the box and $\Phi_{2}\approx \Phi_{1}\approx\Psi_\text{free}/\frac{e^{\I\vect k_c\cdot\vect r_c}}{\sqrt{\Omega}}$. Meanwhile, we can compute the right hand side of \Eq{eq:energy} by applying Gauss's divergence theorem and \Eq{eq:energywavefunction}. Finally we get
\begin{align}
   E_{2}-E_{1}=&\frac{6\hbar^{2}(D^{(0)}_{2}-D^{(0)}_{1})}{M_B \Omega^2}+\frac{\pi^{3}(D_{2}-D_{1})\hbar^{2}}{243\sqrt{3}\Omega^{2}M_B}\nonumber\\
   &\times\sum_{i=1}^3\sum_{j=1}^3\Big(\frac{13}{2}|\vect k_i-\vect k_j|^{4}-5|\vect k_i-\vect k_j|^{2}|\vect k_j-\vect k_r|^{2}\Big),   
\end{align}
where the subscript $r$ is unequal to the subscripts $i$ and $j$. % Specifically, we can let $E_1 = 0$ and $D_1 = 0$, which means the system has no interaction.
Thus we obtain the energy shift caused by the three-body scattering hypervolumes $D^{(0)}$ and $D$:
\begin{align}
   \delta E=&\frac{6\hbar^{2}D^{(0)}}{M_B \Omega^2}+\frac{\pi^{3}\hbar^{2}D}{243\sqrt{3}\Omega^{2}M_B}\nonumber\\
   &\times\sum_{i=1}^3\sum_{j=1}^3\Big(\frac{13}{2}|\vect k_i-\vect k_j|^{4}-5|\vect k_i-\vect k_j|^{2}|\vect k_j-\vect k_r|^{2}\Big).\label{eq:threeenergy}
\end{align}

\section{three-body low-energy scattering in the absence of two-body interactions \label{sec:5}}
The three-body scattering hypervolumes affect the T-matrix element when three bosons collide at low-energy. 
In order to derive the T-matrix element, we notice the following formula for the stationary wave function $\Psi^{E}_{\vect k_1,\vect k_2,\vect k_3}$, with incoming momenta $\hbar\vect k_{1}'/M_B$, $\hbar\vect k_{2}'/M_B$, $\hbar\vect k_{3}'/M_B$, and energy $E=\hbar^2(k_1'^2+k_2'^2+k_3'^2)/2M_B$ in the center-of-mass frame ($\vect k_1'+\vect k_2'+\vect k_3'\equiv0$ and $\vect k_1+\vect k_2+\vect k_3\equiv0$) \cite{tan2008three}:
\begin{widetext}
\begin{equation}
    \Psi^{E}_{\vect k_1,\vect k_2,\vect k_3}=\frac{(2\pi)^{6}}{6}\Big[\sum_{P}\delta(\vect k_1-\vect k_{P1}')\delta(\vect k_2-\vect k_{P2}')\Big]
    +\frac{1}{6}G_{k_{1},k_{2},k_{3}}^{E}T(\vect k_{1}'\vect k_{2}'\vect k_{3}';\vect k_1\vect k_2\vect k_3)
    +\left(\text{terms regular at }k_{1}^{2}+k_{2}^{2}+k_{3}^{2}=2\widetilde{E}\right)\label{eq:Tmatrix}
\end{equation}
\end{widetext}
if the effects of two-body parameters can be neglected,
where $P$ refers to any permutation of 123, and 
\begin{align}
    G^{E}_{k_{1},k_{2},k_{3}}=&\frac{1}{\left(k_{1}^{2}+k_{2}^{2}+k_{3}^{2}\right)/2-\widetilde{E}-\I 0^{+}},\\
    \widetilde{E}=&\frac{k_{1}'^{2}+k_{2}'^{2}+k_{3}'^{2}}{2}.
\end{align}

Furthermore, since the effects of two-body parameters are neglected,~\Eq{eq:111} can be simplified as:
%\begin{equation}
%    \Psi_{L=2}^{M} \approx 4\sqrt{\frac{2\pi}{15}}\bigl[2R^{2}Y_{2}^{M}(\hat{\vect R})+\frac{3}{2}s^{2}Y_{2}^{M}(\hat{\vect s})\bigr]\biggl(1-\frac{D}{B^8}\biggr). \label{eq:threeapprox}
%\end{equation}
\begin{equation}
    \Psi_{L=2}^{M} \approx 4\sqrt{\frac23}\Big[Q_2^M(\vect R)+\frac{3}{4}Q_2^M(\vect s)\Big]\biggl(1-\frac{D}{B^8}\biggr). \label{eq:threeapprox}
\end{equation}
On the other hand, if three identical bosons collide with orbital angular momentum $L=0$, with zero total momentum and zero energy, the most important wave function for
such collision takes the form~\cite{tan2008three}
\begin{equation}
    \Psi_{L=0}\approx 1-\frac{\sqrt{3}D^{(0)}}{8\pi^{3}B^{4}}\label{eq:threeapprox0}
\end{equation}
at large $B$ in the absence of two-body interactions,
where $D^{(0)}$ is the scattering hypervolume defined in Ref.~\cite{tan2008three}.

The Fourier transform of the incoming wave function, namely the first term on the right hand side of~\Eq{eq:Tmatrix}, is $\Psi_{\textbf{in}}=\sum_{P}e^{\I \sum_{i=1}^{3} \vect k_{Pi}'\cdot \vect r_i}/6$. Expanding $\Psi_{\text{in}}$ to the order $B^2$, we get
\begin{align}
    \Psi_{\text{in}}&\simeq 1-\frac{1}{12}\sum_{P}\biggl(\sum_{i=1}^{3}\vect k_{P i}'\cdot r_i\biggr)^{2}\nonumber\\
    &=1-\frac{1}{9}\widetilde{E}B^2\nonumber\\
    &\quad-\frac{1}{18}\sum_{M=-2}^{2}\big[4Q_2^{-M}(\vect p')+3Q_2^{-M}(\vect q')\big]\nonumber\\
    &\mspace{110mu}\times\Big[Q_2^M(\vect R)+\frac34Q_2^M(\vect s)\Big],\label{eq:inexpand}
\end{align}
where $\vect p'=(\vect k_1'-\vect k_2')/2$ and $\vect q'=-\vect k_3'$.
%The first order term disappear since it is proportional to $\vect k_{c}$. 
Comparing the above result with \Eq{eq:threeapprox} and~\Eq{eq:threeapprox0}, we see that the three-body wave function, namely the Fourier transform of $\Psi^E_{\vect k_1,\vect k_2,\vect k_3}$, can be approximated at $r_e\ll B\ll \lambda$, where $\lambda\sim 1/k_1'\sim 1/k_2'\sim 1/k_3'$:
\begin{align}
    \Psi^{E}(\vect R, \vect s)\approx & 1-\frac{\sqrt{3}D^{(0)}}{8\pi^{3}B^{4}}-\frac{1}{9}\widetilde{E}B^2\nonumber\\&
    -\frac{1}{18}\sum_{M=-2}^{2}
    \left[4Q_{2}^{-M}(\mathbf{p'})+3Q_{2}^{-M}(\mathbf{q'})\right]\nonumber\\
    &\times\Big[Q^{M}_{2}(\vect R)+\frac{3}{4}Q_{2}^{M}(\vect s)\Big]\Big(1-\frac{D}{B^8}\Big).
\end{align}
Taking the inverse Fourier transformation of the above equation, we get
\begin{align}
    \Psi^{E}_{\vect k_1,\vect k_2,\vect k_3}\approx &(2\pi)^{6}\delta(\vect p)\delta(\vect q)-\frac{D^{(0)}}{p^{2}+\frac{3}{4}q^{2}}\nonumber\\&
    +\frac{(2\pi)^{6}}{36}\bigl(4p'^2+3q'^3\bigr)\left(\nabla^{2}_{\vect q}+\frac{3}{4}\nabla^{2}_{\vect p}\right)\delta(\vect p)\delta(\vect q)\nonumber\\&
    +\sum_{M=-2}^{2}\frac{(2\pi)^{6}}{18}\biggl[4Q^{-M}_{2}(\vect p')+3Q^{-M}_{2}(\vect q')\biggr] \nonumber\\&
    \times\biggl[Q_{2}^{M}(\nabla_{\vect q})+\frac{3}{4}Q_{2}^{M}(\nabla_{\vect p})\biggr]\delta(\vect p)\delta(\vect q) \nonumber\\&
    -\frac{\pi^{3}D}{54\sqrt{3}}\frac{f(\vect p,\vect q,\vect p',\vect q')}{p^{2}+\frac{3}{4}q^{2}},\label{eq:Teigen}
\end{align}
where $\mathbf{p}=(\vect k_1-\vect k_2)/2$ and $\mathbf{q}=-\vect k_3$, and
\begin{align}
    f(\mathbf{p},\mathbf{q},\mathbf{p'},\mathbf{q'})=\frac{1}{3}\sum_{M=-2}^2 &\left[4Q_{2}^{M}(\mathbf{p'})+3Q_{2}^{M}(\mathbf{q'})\right]\nonumber\\&
    \times
    \left[4Q_{2}^{-M}(\mathbf{p})+3Q_{2}^{-M}(\mathbf{q})\right]. \label{eq:f}
\end{align}
Comparing~\Eq{eq:Teigen} with \Eq{eq:Tmatrix}, we obtain an approximate result for the T-matrix element:
\begin{equation}\label{Tmatrixresult}
    T(\vect k_{1}'\vect k_{2}'\vect k_{3}';\vect k_1\vect k_2\vect k_3)\approx-6D^{(0)}-\frac{\pi^{3}D}{9\sqrt{3}}f(\mathbf{p},\mathbf{q},\mathbf{p'},\mathbf{q'}).
\end{equation}

\section{Construction of the effective Hamiltonian \label{sec:6}}
In order to study the effects of the three-body scattering hypervolumes on the many-body physics in the low-density and low-temperature regime (with bosonic number density $n\to0$ and absolute temperature $T\to0$), we can construct an effective Hamiltonian containing a kinetic energy term $\hat{K}$ and some effective three-body term $\hat{V}$:
\begin{equation}\label{Hamiltonian}
H=\hat{K}+\hat{V},
\end{equation}
where
\begin{equation}
\hat{K}=\sum_{\vect k}\frac{\hbar^{2}k^{2}}{2M_{B}}a^{\dagger}_{\vect k}a_{\vect k},
\end{equation}
\begin{align}
    \hat{V}=&\frac{1}{6\Omega^{2}}\sum_{\vect k_1, \vect k_2, \vect k_3}\sum_{\vect k'_1, \vect k'_2, \vect k'_3}g(\vect k'_1, \vect k'_2, \vect k'_3; \vect k_1, \vect k_2, \vect k_3)\nonumber\\&
    \times  \delta_{\vect k_1+\vect k_2+\vect k_3,\vect k_1'+\vect k_2'+\vect k_3'}a^{\dagger}_{\vect k_1}a^{\dagger}_{\vect k_2}a^{\dagger}_{\vect k_3}a_{\vect k'_1}a_{\vect k'_2}a_{\vect k'_3},\label{eq:hatV}
\end{align}
and $a_\vect k$ is the boson annihilation operator for the single-particle state with momentum $\hbar\vect k$.
Now consider a three-body state describing three bosons colliding with incoming linear momenta $\hbar\vect k_1'$, $\hbar\vect k_2'$, and $\hbar\vect k_3'$ satisfing $\vect k_3'=-\vect k_1'-\vect k_2'$:
\begin{equation}
    \ket{\vect k'_1, \vect k'_2, \vect k'_3}=\sum_{\vect k_1, \vect k_2}\Psi_{\vect k_1, \vect k_2, \vect k_3}a_{\vect k_1}^{\dagger}a_{\vect k_2}^{\dagger}a_{\vect k_3}^{\dagger}\ket{0},\label{eq:threebodywavefunction}
\end{equation}
where $\vect k_3\equiv-\vect k_1-\vect k_2$, $\Psi_{\vect k_1,\vect k_2,\vect k_3}$ is the three-body wave function,
and $\ket{0}$ is the normalized vacuum state. Requiring that $\ket{\vect k'_1, \vect k'_2, \vect k'_3}$ be the eigenstate of
$H$, and assuming that $\Psi_{\vect k_1\vect k_2\vect k_3}$ may be approximated by the function $\Psi^E_{\vect k_1\vect k_2\vect k_3}$ studied in the last section in the large-box limit ($\Omega\to\infty$), we find that
\begin{align}
g(\vect k'_1, \vect k'_2, \vect k'_3; \vect k_1, \vect k_2, \vect k_3)\approx &-\frac{\hbar^2}{6M_B}T(\vect k_1'\vect k_2'\vect k_3';\vect k_1\vect k_2\vect k_3). \label{eq:gexpression}
\end{align} Substituting \Eq{Tmatrixresult} into \Eq{eq:gexpression}, we get
\begin{align}\label{eq:gkkkkkk}
    g(\vect k'_1, \vect k'_2, \vect k'_3; \vect k_1, \vect k_2, \vect k_3)\approx &\frac{\hbar^{2}}{M_{B}}\left[D^{(0)}+\frac{\pi^{3}D}{54\sqrt{3}}f(\mathbf{p},\mathbf{q},\mathbf{p'},\mathbf{q'})\right].
\end{align}

We may compute the first-order effects of the three-body scattering hypervolumes on the three-body energy eigenvalue in the large periodic box by computing the expectation value of  $\hat{V}$ in the unperturbed state. Assuming that it is a symmetric three-body state which $\vect k_1$, $\vect k_2$, and $\vect k_3$ are all different, we set a state $\ket{\text{free}}=a_{\vect k_1}^{\dagger}a_{\vect k_2}^{\dagger}a_{\vect k_3}^{\dagger}\ket{0}$. The three-body energy shift is
\begin{align}
\delta E\approx\bra{\text{free}}\hat{V}\ket{\text{free}},
\end{align}
to first order in the three-body scattering hypervolumes. Substituting \Eq{eq:hatV} and using Eqs.~\eqref{eq:gkkkkkk} and \eqref{eq:f}, we find that $\delta E$ is the same as \Eq{eq:threeenergy}.

\section{Energy shift in the thermodynamic limit \label{sec:7}}
In this section we study the effects of the three-body scattering hypervolumes on the energy of a dilute Bose gas.

We consider a dilute homogeneous Bose gas containing $N$ bosons on average, occupying volume $\Omega$,
and consider the thermodynamic limit with a fixed number density $n=N/\Omega$.
We assume that the bosons have short-range rotationally-invariant and translationally-invariant interactions, with range $r_e$,
and that the number density $n$ is sufficiently small such that $nr_e^3\ll1$.
If we start from an unperturbed mixed state with density operator $\rho_\text{unperturbed}$ describing an ideal Bose gas with absolute temperature $T$ and chemical potential $\mu$ in the grand canonical ensemble,
and introduce the interactions adiabatically such that we obtain three-body scattering hypervolumes $D^{(0)}$ and $D$,
the shift of the many-body energy expectation $E$ due to the three-body scattering hypervolumes may be approximately calculated:
\begin{equation}\label{DeltaE}
\Delta E\approx\mathrm{Tr}(\rho_\text{unperturbed}\hat{V}),
\end{equation}
where $\hat{V}$ is defined in the last section, and $\mathrm{Tr}$ is the trace.

\subsection{$T>T_c$}
If the temperature $T$ is above the Bose-Einstein condensation (BEC) critical temperature $T_c$, the mean occupation of the zero momentum state, namely the expectation value of $a_\vect 0^\dagger a_\vect 0$, is very small compared to $N$.
Thus, we can calculate the expectation value $\mathrm{Tr}(\rho_\text{unperturbed}\hat{V})$ by using the Wick's theorem
\begin{align}
&\mathrm{Tr}(\rho_\text{unperturbed}a_{\vect k_1}^\dagger a_{\vect k_2}^\dagger a_{\vect k_3}^\dagger a_{\vect k_1'}a_{\vect k_2'}a_{\vect k_3'})\nonumber\\
&=\sum_{P}n_{\vect k_1}n_{\vect k_2}n_{\vect k_3}\delta_{\vect k_{1}\vect k_{P1}'}\delta_{\vect k_{2}\vect k_{P2}'}\delta_{\vect k_{3}\vect k_{P3}'},\label{aaaaaacase1}
\end{align}
where $P$ is any of the six permutations of $123$,
\begin{equation}
n_{\vect k}=\frac{1}{\exp[\beta(\epsilon_\vect k-\mu)]-1}
\end{equation}
is the Bose-Einstein distribution, $\beta=1/k_BT$, $k_B$ is the Boltzmann constant, and $\epsilon_\vect k=\hbar^2k^2/2M_B$
is the kinetic energy of a boson with momentum $\hbar\vect k$.
The result is:
%\begin{align}\label{DeltaET>Tc}
%\frac{\Delta E}{\Omega}=\frac{\hbar^{2}D^{(0)}n^{3}}{M_{B}}+&\frac{10 D M_B^{4}n}{2187\sqrt{3}\hbar^{8}\beta^{5}}\biggl\{  \big[g_{\frac{5}{2}}(e^{\beta \mu})\big]^{2}\nonumber\\&
%+ 16\pi^{\frac{3}{2}}\left(\frac{ T_{d}}{T}\right)^{\frac{3}{2}}g_{\frac{7}{2}}(e^{\beta \mu})\biggr\},
%\end{align}
\begin{align}\label{DeltaET>Tc}
\frac{\Delta E}{\Omega}=\frac{\hbar^{2}D^{(0)}n^{3}}{M_{B}}+&\frac{5D \hbar^{2}n^{13/3}}{34992\sqrt{3}M_{B}}\Big\{ \widetilde{T}^{5} \big[g_{\frac{5}{2}}(e^{\beta \mu})\big]^{2}\nonumber\\&
+ 16\pi^{\frac{3}{2}}\widetilde{T}^{7/2}g_{\frac{7}{2}}(e^{\beta \mu})\Big\},
\end{align}
where
\begin{equation}
\widetilde{T}\equiv T/T_d,
\end{equation}
\begin{equation}
T_{d}\equiv\frac{\hbar^{2}n^{\frac{2}{3}}}{2 M_Bk_{B}}
\end{equation}
is the quantum degeneracy temperature, and $g_{\nu}\left(z\right)$ is defined as \cite{Pethick2008}
\begin{equation}
    g_{\nu}(z)=\frac{1}{\Gamma(\nu)}\int_{0}^{\infty}dx \frac{x^{\nu -1}}{z^{-1}e^{x}-1},
\end{equation}
where $\Gamma(\nu)$ is the Gamma function.
The chemical potential satisfies the following equation:
\begin{equation}
    n=\left(\frac{M_Bk_BT}{2\pi \hbar^{2}}\right)^{\frac{3}{2}}g_{\frac{3}{2}}(e^{\beta \mu}).
\end{equation}

When $T=T_c$, \Eq{DeltaET>Tc} is reduced to
\begin{equation}
\frac{\Delta E}{\Omega}=\frac{\hbar^2D^{(0)}n^3}{M_B}+J\frac{\hbar^2Dn^{13/3}}{M_B},
\end{equation}
where
\begin{equation}
J=\frac{320\pi^{5}(z_{5}^{2}+2z_{3}z_{7})}{2187\sqrt{3}z_{3}^{10/3}}\approx8.09281, \label{eq:J}
\end{equation}
and we have defined a short-hand notation
\begin{equation}
z_n\equiv\zeta(n/2),
\end{equation}
where $\zeta\left(x\right)$ is the Riemann-Zeta function.

When $T$ is above $T_c$ and comparable to $T_c$, the second term on the right hand side of \Eq{DeltaET>Tc} is on the order of $\hbar^2Dn^{13/3}/M_B$. The ratio between the second term and the first term on the right hand side of \Eq{DeltaET>Tc} is on the order of $Dn^{4/3}/D^{(0)}$. This ratio is normally much smaller than $1$ at $nr_e^3\ll1$, suggesting that the effect of the three-body $L=2$
scattering hypervolume $D$ is much smaller than that of the $L=0$ scattering hypervolume $D^{(0)}$.
However, if the system is near an $L=2$ three-body resonance, such that $D$ is anomalously large, then the second term on the right hand side of \Eq{DeltaET>Tc} may be more significant.

When $T_c\ll T\ll T_e$ [where $T_e=\hbar^2/(2M_B r_e^2k_B)$ is the temperature scale at which the thermal de Broglie wave lengths of the bosons are comparable to the range of the interaction $r_e$], \Eq{DeltaET>Tc} is reduced to
\begin{equation}
\frac{\Delta E}{\Omega}\approx\frac{\hbar^2D^{(0)}n^3}{M_B}+\frac{80\pi^3M_BDn^3(k_BT)^2}{729\sqrt3\,\hbar^2}.
\end{equation}

\subsection{$T< T_c$}
If $T<T_c$, we can approximate the boson operator as \cite{pitaevskii2016}
\begin{equation}
    a_{\mathbf{k}}=\sqrt{N_{0}}\delta_{\mathbf{k},0}+a_{\mathbf{k}\neq 0},
\end{equation}
where $N_{0}$ is the number of bosons in the condensate. To calculate $\mathrm{Tr}(\rho_\text{unperturbed}a_{\vect k_1}^\dagger a_{\vect k_2}^\dagger a_{\vect k_3}^\dagger a_{\vect k_1'}a_{\vect k_2'}a_{\vect k_3'})$ approximately,
we need to consider four cases \cite{dogra2019}:
\begin{itemize}
\item Case 1: $\vect k_1$, $\vect k_2$, and $\vect k_3$ are all nonzero, in which \Eq{aaaaaacase1} is applicable, with $\mu$ set to $0$.
\item Case 2: one of the three wave vectors $\vect k_1$, $\vect k_2$, and $\vect k_3$ is zero. For instance, if $\vect k_1=\vect 0$ but $\vect k_2$ and $\vect k_3$ are both nonzero,
then
\begin{align}
&\mathrm{Tr}(\rho_\text{unperturbed}a_{\vect k_1}^\dagger a_{\vect k_2}^\dagger a_{\vect k_3}^\dagger a_{\vect k_1'}a_{\vect k_2'}a_{\vect k_3'})\nonumber\\
&=\sum_{P}N_0n_{\vect k_2}n_{\vect k_3}\delta_{\vect 0,\vect k_{P1}'}\delta_{\vect k_{2}\vect k_{P2}'}\delta_{\vect k_{3}\vect k_{P3}'}.\label{aaaaaacase2}
\end{align}
\item Case 3: two of the three wave vectors $\vect k_1$, $\vect k_2$, $\vect k_3$ are zero. For instance, if $\vect k_1=\vect k_2=\vect 0$ but $\vect k_3\ne\vect 0$, then
\begin{align}
&\mathrm{Tr}(\rho_\text{unperturbed}a_{\vect k_1}^\dagger a_{\vect k_2}^\dagger a_{\vect k_3}^\dagger a_{\vect k_1'}a_{\vect k_2'}a_{\vect k_3'})\nonumber\\
&=N_0^2n_{\vect k_3}\big(\delta_{\vect 0,\vect k_1'}\delta_{\vect 0,\vect k_2'}\delta_{\vect k_3,\vect k_3'}
+\delta_{\vect 0,\vect k_2'}\delta_{\vect 0,\vect k_3'}\delta_{\vect k_3,\vect k_1'}\nonumber\\
&\mspace{80mu}+\delta_{\vect 0,\vect k_3'}\delta_{\vect 0,\vect k_1'}\delta_{\vect k_3,\vect k_2'}).\label{aaaaaacase3}
\end{align}
\item Case 4: $\vect k_1=\vect k_2=\vect k_3=\vect 0$, so that
\begin{align}
&\mathrm{Tr}(\rho_\text{unperturbed}a_{\vect k_1}^\dagger a_{\vect k_2}^\dagger a_{\vect k_3}^\dagger a_{\vect k_1'}a_{\vect k_2'}a_{\vect k_3'})\nonumber\\
&=N_0^3\delta_{\vect 0,\vect k_1'}\delta_{\vect 0,\vect k_2'}\delta_{\vect 0,\vect k_3'}.\label{aaaaaacase4}
\end{align}
\end{itemize}
Then we derive from \Eq{DeltaE} the following result:
\begin{align}
    \frac{\Delta E}{\Omega}&=\frac{D^{(0)}\hbar^{2}}{6M_{B}}\left(6n'^{3}+18n'^{2}n_{0}+9n_{0}^2 n'+n_{0}^{3}\right)\nonumber\\&
    \quad+\frac{5D \hbar^{2}n^{13/3}}{34992\sqrt{3}M_{B}}\nonumber\\&
    \quad\quad\times\Big(8\pi^{3/2}z_{7}\frac{n_{0}^{2}+4n_{0}n'+2n'^{2}}{n^{2}}\widetilde{T}^{7/2}+z_5^2\widetilde{T}^{5}\Big)\nonumber\\&
    =\frac{D^{(0)} \hbar^2 n^3}{6 M_{B}} \Xi_{0}(\widetilde{T})+\frac{5 D \hbar^2 n^{13/3}}{279936 \sqrt{3} \pi ^{\frac{3}{2}} M_{B}}\Xi_{2}(\widetilde{T}),
\end{align}
\begin{equation}
    \Xi_{0}(\widetilde{T})\equiv 1+\frac{3z_3}{4\pi^{3/2}}\widetilde{T}^{3/2}+\frac{3z_3^2}{64\pi^3}\widetilde{T}^3-\frac{z_3^3}{128\pi^{9/2}}\widetilde{T}^{9/2},\label{Xi0}
\end{equation}
\begin{equation}
    \Xi_{2}(\widetilde{T})\equiv 64 \pi ^3  z_{7}\widetilde{T}^{7/2}+8 \pi ^{3/2} \left(z_{5}^2+2 z_{3} z_{7}\right) \widetilde{T}^{5}- z_{3}^2 z_{7}\widetilde{T}^{13/2},\label{Xi2}
\end{equation}
where $n_{0}=N_0/\Omega$ is the number density of bosons in the condensate, and $n'=(N-N_0)/\Omega$ is the number density of bosons outside of the condensate.
If $T\ll T_c$,
\begin{align}
    \frac{\Delta E}{\Omega} \approx &\frac{\hbar^2D^{(0)}  n^3}{6 M_{B}}\Big(1+\frac{3 z_{3}\widetilde{T}^{3/2}}{4 \pi ^{3/2}} \Big)
    +\frac{5 \pi ^{3/2} z_{7}\hbar^2 Dn^{13/3}}{4374 \sqrt{3} M_{B}}\widetilde{T}^{7/2}.
\end{align}
If $T= 0$, $\Delta E= \frac{D^{(0)} \hbar^{2}\Omega}{6 M_{B}}n^{3}$, which is consistent with Ref.~\cite{tan2008three}. If $T$ is comparable to $T_{c}$, the ratio between the $L=2$ term and the $L=0$ term is on the order of $Dn^{4/3}/D^{(0)}$.

\begin{figure}
    \includegraphics[width=1\linewidth]{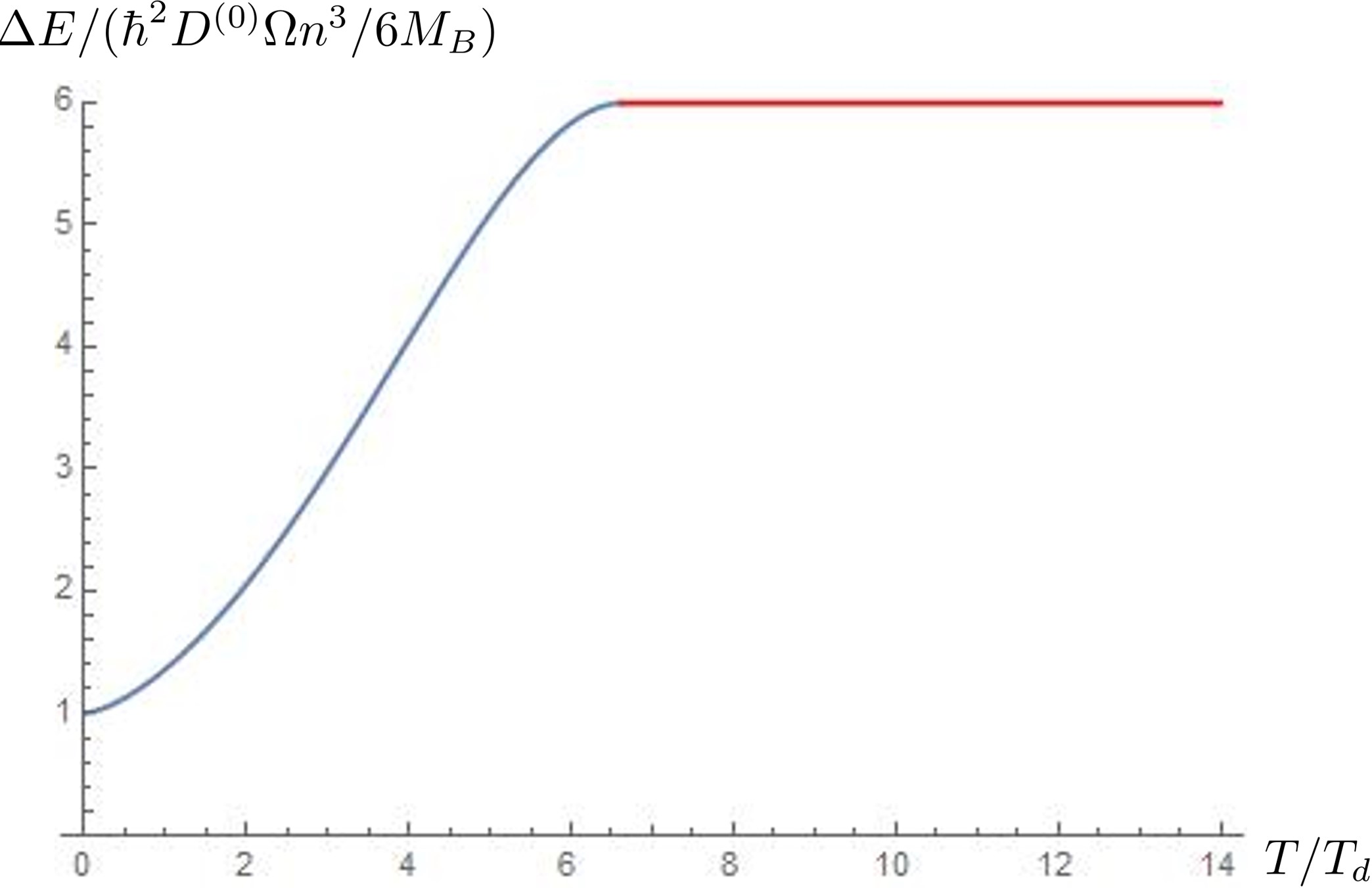}
    \caption{The energy shift of the dilute Bose gas caused by the adiabatic introduction of the three-body $L=0$ scattering hypervolume $D^{(0)}$ vs. the reduced temperature $T/T_d$.  The red line shows the energy shift above the critical temperature, and the blue curve shows the energy shift below the critical temperature.}
    \label{fig:a}
\end{figure}

\begin{figure}
    \includegraphics[width=1\linewidth]{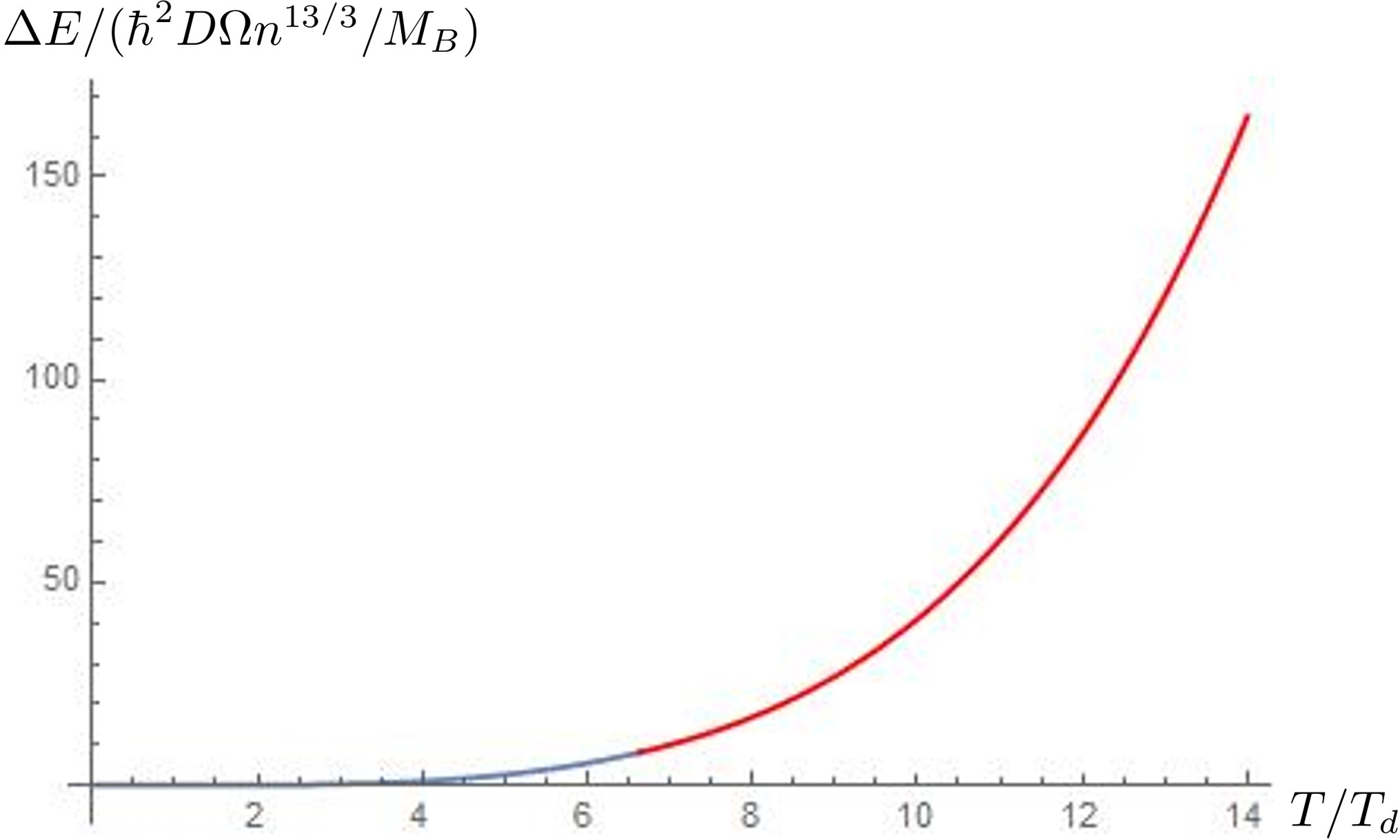}
    \caption{The energy shift of the dilute Bose gas by the adiabatic introduction of the three-body $L=2$ scattering hypervolume $D$ vs. the reduced temperature $T/T_d$.  The red curve shows the energy shift above the critical temperature, and the blue curve shows the energy shift below the critical temperature.}
    \label{fig:b}
\end{figure}

In Fig.~\ref{fig:a} we plot the shift of energy due to the three-body $L=0$ scattering hypervolume $D^{(0)}$.
In Fig.~\ref{fig:b} we plot the shift of energy due to the three-body $L=2$ scattering hypervolume $D$.

\section{three-body recombination rate \label{sec:8}}
If the two-body interaction does \emph{not} support any two-body bound state,
the three-body collisions are purely elastic, and the three-body scattering hypervolumes $D^{(0)}$ and $D$ are real \cite{tan2008three}.
If the interaction supports at least one two-body bound state, the bound pair may be formed during three-body collisions,
and $D_{0}$ and $D$ will in general contain negative imaginary parts \cite{shinazhu2017}, and in the Bose gas we have three-body
recombination processes, in which two bosons will form a dimer (bound pair) and the energy is released in the form of kinetic energy such
that the dimer and the third boson may leave the trap \cite{moerdijk1996,fedichev1996,nielsen1999,esry1999,braaten2001,hammer2013,Mirahmadi2022}, so that the number density $n$ of the remaining low-energy bosons decreases:
\begin{equation}
    \frac{dn}{dt}=-L_{3}n^{3} ,\label{eq:trdef}
\end{equation}
where $t$ is time, and $L_{3}$ is the three-body recombination rate constant. Different three-body recombination channels contribute to  $L_3$. The three-body $L=0$ scattering hypervolume defined in Ref.~\cite{tan2008three} is usually the dominant parameter in $L_3$ at low temperatures ($T\ll T_e$), while the three-body $L=2$ scattering hypervolume $D$ defined in this paper will contribute a correction to $L_{3}$.

If the system undergoes three-body recombination, the norm square of the wave function decays with a factor $e^{-2|\mathrm{Im}E|\Delta t/\hbar}$ within a short time $\Delta t$. This is the probability that three-body recombination does not occur. Thus, the probability that three-body recombination happens can be approximated as $2|\mathrm{Im}E|\Delta t/\hbar$. Since the system will lose three low-energy bosons after each three-body recombination event, the change of the number of low-energy bosons during a short time $dt$ is
\begin{equation}
    dN=-3\frac{2dt}{\hbar}|\mathrm{Im}E|. \label{eq:NLOss}
\end{equation}
Using \Eq{eq:trdef}, \Eq{eq:NLOss}, and the results of the last section, we find that if $T>T_c$,
\begin{align}
    L_{3}=\frac{6\hbar|\text{Im}D^{(0)}|}{M_{B}}+&\frac{5\hbar|\text{Im}D| n^{4/3}}{5832\sqrt{3}M_{B}}\Big\{ \widetilde{T}^{5} \big[g_{\frac{5}{2}}(e^{\beta \mu})\big]^{2}\nonumber\\&
    + 16\pi^{\frac{3}{2}}\widetilde{T}^{7/2}g_{\frac{7}{2}}(e^{\beta \mu})\Big\}.\label{eq:trr>}
\end{align}
If $T_c\ll T\ll T_e$, the above result may be approximated as
\begin{equation}
     L_{3}\approx \frac{6\hbar|\text{Im}D^{(0)}|}{M_{B}}+ \frac{160 \pi^{3}M_{B}|\mathrm{Im}D|}{243\sqrt{3}\hbar^{3}}(k_{B}T)^{2} .
\end{equation}
At $T=T_c$, we get
\begin{equation}
    L_{3}=\frac{6\hbar|\text{Im}D^{(0)}|}{M_{B}}+\frac{6J\hbar|\mathrm{Im}D|}{M_B}n^{\frac{4}{3}},\label{eq:recombinationTc}
\end{equation}
where $J$ is defined in~\Eq{eq:J}. 
If $T$ is above $T_{c}$ and comparable to $T_{c}$, the ratio between the contribution of $D$ to $L_3$ and that of $D^{(0)}$ to $L_3$
on the right hand side of \Eq{eq:trr>} is on the order of $|\text{Im}D|n^{4/3}/|\text{Im}D^{(0)}|$. This ratio is normally much smaller than 1 for dilute Bose gases, so that the effect of $D^{(0)}$ is normally dominant. However, if the system is near a three-body $L=2$ resonance, such that $D$ is anomalously large, then the contribution due to $D$ may be more important.

If $T< T_{c}$, we get
\begin{align}
    L_{3}=\frac{\hbar|\text{Im}D^{(0)}|}{ M_{B}} \Xi_{0}(\widetilde{T})+\frac{5\hbar |\text{Im}D| n^{4/3}}{46656 \sqrt{3} \pi ^{\frac{3}{2}} M_{B}}\Xi_{2}(\widetilde{T}),
\end{align}
where the functions $\Xi_0(\widetilde{T})$ and $\Xi_2(\widetilde{T})$ are defined in Eqs.~\eqref{Xi0} and \eqref{Xi2}.
If $T\ll T_{c}$, the above result may be approximated as
\begin{align}
    L_{3}\approx &\frac{\hbar|\text{Im}D^{(0)|}}{ M_{B}}\Big(1+\frac{3 z_{3} \widetilde{T}^{3/2}}{4 \pi ^{3/2}}\Big)\nonumber\\&
    +\frac{5z_{7} \pi ^{3/2}\hbar |\text{Im}D| n^{4/3} }{729 \sqrt{3} M_{B}}\widetilde{T}^{7/2}.
\end{align}
If $T= 0$, $L_{3}(T)= \frac{\hbar |\text{Im}D^{(0)}|}{M_B}$, which is consistent with Ref.~\cite{shinazhu2017}.

\section{summary and discussion}
In this paper, we have studied three identical bosons with short-range interactions colliding with total orbital angular momentum $L=2$. We derived the 111-expansion and 21-expansion of the wave function at zero collision energy and defined the three-body $L=2$ scattering hypervolume $D$ whose dimension is length to the 8th power. $D$ appears at the order $B^{-6}$ in the 111-expansion and appears at the order $R^{-6}$ in the 21-expansion. If there is at least a two-body bound state, $D$ will in general become complex. Both $D^{(0)}$ (which is defined in Ref.~\cite{tan2008three}) and $D$ play important roles in the three-body T-matrix element for low-energy collisions. 

The values of $D^{(0)}$ and $D$ can be calculated by solving the three-body Schrödinger equation numerically in general.
For weak interactions, we have derived an approximate formula of $D$ by using the Born expansion.
If the absolute value of the two-body scattering length $a$ is much larger than the range of interaction, there is Efimov effect
for $L=0$ collisions of three identical bosons \cite{efimov1970}, which should lead to a universal formula for $D^{(0)}$ in terms of 
$a$ and the Efimov scale \footnote{Shangguo Zhu, private discussion}, related to the three-body recombination rate \cite{nielsen1999,esry1999,braaten2000} and the energy \cite{braaten2002} of dilute Bose gases with large scattering length.
However, there should be no Efimov effect for $L=2$ collisions of three identical bosons in three dimensions \cite{efimov1970},
and we expect that there should be a universal formula for $D$ in terms of $a$ if we approximate the interactions as contact interactions.
%\begin{equation}
%D_\Lambda=\Big[\eta\big(\mathrm{sign}(a)\big)-\frac{3}{32\pi}\sqrt\frac{15}{2}J_1\ln\frac{\Lambda}{|a|}\Big] a^8,
%\end{equation}
%which depends on the arbitrary length scale $\Lambda$ in the $111$-expansion \Eq{eq:111} (in such a way that the resultant $111$-expansion does not depend on $\Lambda$) and the sign of $a$.

The three-body scattering hypervolumes $D^{(0)}$ and $D$ are important not only in the few-body physics but also in the many-body physics. We studied the energy shift of three bosons in a large periodic box due to the adiabatic introduction of interactions and found the energy shift due to $D^{(0)}$ and $D$. 
We also studied the energy shift of the dilute Bose gas in the thermodynamic limit due to the adiabatic introduction of interactions and found the energy shift due to $D^{(0)}$ and $D$, from which we also computed the three-body recombination rate constant in terms of the imaginary parts of $D^{(0)}$ and $D$.
In general, the contributions of $D$ to the properties of the dilute Bose gas are much smaller than those of $D^{(0)}$. However, if the system is near a three-body $L=2$ resonance, $D$ may be anomalously large
and the effects of $D$ may be more important.

\begin{acknowledgments}
This work was supported by the National Natural Science Foundation of China Grant No.92365202 and the 
National Key R\&D Program of China (Grants No. 2019YFA0308403 and No. 2021YFA1400902). We thank Junjie
Liang and Jiansen Zhang for fruitful discussions.
\end{acknowledgments}

\appendix

\section{111-EXPANSION AND 21-EXPANSION: COEFFICIENT LIST\label{appendix:coefficients}}

\begin{equation}
c_{0,0}^{0}(\vect R)=\frac{64\pi-144\sqrt{3}}{3\pi}\sqrt{\frac{2\pi}{15}}Y_{2}^{M}(\mathbf{\hat{R}}),
\end{equation}
\begin{equation}
c_{2,M}^{0}(\vect R)=30\sqrt{6\pi}Y_{0}^{0}(\mathbf{\hat{R}}),
\end{equation}
\begin{equation}
c_{0,0}^{-1}(\vect R)=\frac{2a^{2} A_{1}}{3\pi R}\sqrt{\frac{2\pi}{15}}Y^{M}_{2}(\mathbf{\hat{R}}),
\end{equation}
\begin{align}
c_{2,m}^{-1}(\vect R)=&-\frac{12\pi a}{7R}\sqrt{\frac{21}{4\pi}}C^{2,M}_{4,M-m;2,m}Y^{M-m}_{4}(\mathbf{\hat{R}})\nonumber\\&
+\frac{96\pi a}{R}\sqrt{\frac{5}{84\pi}}C^{2,M}_{2,M-m;2,m}Y^{M-m}_{2}(\mathbf{\hat{R}})\nonumber\\&
-\frac{336\pi a}{R}\sqrt{\frac{1}{6\pi}}Y^{0}_{0}(\mathbf{\hat{R}})\delta_{M,m},
\end{align}
\begin{equation}
c_{0,0}^{-2}(\vect R)=\frac{\left(8w_{2}A_{1}-108\pi b_{2}r_{s}\right)a^{3}}{9\pi^{2}R^{2}}\sqrt{\frac{2\pi}{15}}Y_{2}^{M}(\mathbf{\hat{R}}),
\end{equation}
\begin{align}
c_{2,m}^{-2}(\vect R)=&-\frac{32b_{2}a^{2}}{\sqrt{3}R^{2}}\sqrt{\frac{5}{28\pi}}C^{2,M}_{2,M-m;2,m}Y_{2}^{M-m}(\mathbf{\hat{R}})\nonumber\\&
-\frac{384d_{2}a^{2}}{R^{2}}\sqrt{\frac{1}{28\pi}}C^{2,M}_{4,M-m;2,m}Y_{4}^{M-m}(\mathbf{\hat{R}})\nonumber\\&
+\frac{8c_{2}a^{2}}{R^{2}}\sqrt{\frac{1}{6\pi}}Y_{0}^{0}(\mathbf{\hat{R}})\delta_{M,m},
\end{align}
\begin{align}
c_{4,m}^{-3}(\vect R)=&-\frac{3924a}{R^{3}}\sqrt{\frac{3\pi}{35}}C^{2,M}_{2,M-m;4,m}Y_{2}^{M-m}(\mathbf{\hat{R}})\nonumber\\&
+\frac{900a}{R^{3}}\sqrt{\frac{6\pi}{77}}C^{2,M}_{4,M-m;4,m}Y_{4}^{M-m}(\mathbf{\hat{R}})\nonumber\\&
-\frac{315a}{2R^{3}}\sqrt{\frac{3\pi}{22}}C^{2,M}_{6,M-m;4,m}Y_{6}^{M-m}(\mathbf{\hat{R}}),
\end{align}
\begin{align}
c_{2,m}^{-3}(\vect R)=&-\frac{4A_{3}}{R^{3}}\sqrt{\frac{5}{21\pi}}C^{2,M}_{2,M-m;2,m}Y_{2}^{M-m}(\mathbf{\hat{R}})\nonumber\\&
+\frac{9B_{3}}{\sqrt{7\pi}R^{3}}C^{2,M}_{4,M-m;2,m}Y_{4}^{M-m}(\mathbf{\hat{R}}),
\end{align}
\begin{equation}
c_{2,M}^{-3}(\vect R)=\frac{32b_{2}a^{3}}{R^{3}}\sqrt{\frac{2}{3\pi}}Y_{0}^{0}(\mathbf{\hat{R}}),
\end{equation}
\begin{equation}
c_{0,0}^{-3}(\vect R)=-\frac{4\eta}{a R^{3}}\sqrt{\frac{2\pi}{15}}Y_{2}^{M}(\mathbf{\hat{R}}),
\end{equation}
\begin{align}
c_{4,m}^{-4}(\vect R)=&\frac{48w_{4}a^{2}}{R^{4}}\sqrt{\frac{3}{35\pi}}C^{2,M}_{2,M-m;4,m}Y_{2}^{M-m}(\mathbf{\hat{R}})\nonumber\\&
-\frac{96b_{4}a^{2}}{R^{4}}\sqrt{\frac{6}{77\pi}}C^{2,M}_{4,M-m;4,m}Y_{4}^{M-m}(\mathbf{\hat{R}})\nonumber\\&
+\frac{1536c_{4}a^{2}}{R^{4}}\sqrt{\frac{2}{11\pi}}C^{2,M}_{6,M-m;4,m}Y_{6}^{M-m}(\mathbf{\hat{R}}),
\end{align}
\begin{align}
c_{2,m}^{-4}(\vect R)=&-\frac{A_{4}}{3\sqrt{7}\pi^{\frac{3}{2}}R^{4}}C^{2,M}_{4,M-m;2,m}Y_{4}^{M-m}(\mathbf{\hat{R}})\nonumber\\&
+\frac{B_{4}}{9\pi R^{4}}\sqrt{\frac{5}{7\pi}}C^{2,M}_{2,M-m;2,m}Y_{2}^{M-m}(\mathbf{\hat{R}})\nonumber\\&
-\frac{C_{4}}{9\pi R^{4}}\sqrt{\frac{2}{\pi}}Y_{0}^{0}(\mathbf{\hat{R}})\delta_{M,m},
\end{align}

\begin{equation}
c_{0,0}^{-4}(\vect R)=\frac{4\alpha}{a}\sqrt{\frac{2\pi}{15}}Y_{2}^{M}(\mathbf{\hat{R}}),
\end{equation}
\begin{align}
c_{6,m}^{-5}(\vect R)=&-\frac{693a\pi}{4R^{5}}\sqrt{\frac{273}{2\pi}}C^{2,M}_{8,M-m;6,m}Y_{8}^{M-m}(\mathbf{\hat{R}})\nonumber\\&
+\frac{5733a\pi}{R^{5}}\sqrt{\frac{21}{11\pi}}C^{2,M}_{6,M-m;6,m}Y_{6}^{M-m}(\mathbf{\hat{R}})\nonumber\\&
-\frac{27405a\pi}{R^{5}}\sqrt{\frac{39}{22\pi}}C^{2,M}_{4,M-m;6,m}Y_{4}^{M-m}(\mathbf{\hat{R}}),
\end{align}
\begin{align}
c_{4,m}^{-5}(\vect R)=&\frac{48b_{2}a^{3}}{R^{5}}\sqrt{\frac{105}{\pi}}C^{2,M}_{2,M-m;4,m}Y_{2}^{M-m}(\mathbf{\hat{R}})\nonumber\\&
-\frac{A_{5}}{R^{5}}\sqrt{\frac{2}{11\pi}}C^{2,M}_{6,M-m;4,m}Y_{6}^{M-m}(\mathbf{\hat{R}})\nonumber\\&
-\frac{B_{5}}{R^{5}}\sqrt{\frac{42}{11\pi}}C^{2,M}_{4,M-m;4,m}Y_{4}^{M-m}(\mathbf{\hat{R}}),
\end{align}
\begin{align}
c_{2,m}^{-5}(\vect R)=&-\frac{C_{5}}{R^{5}}C^{2,M}_{4,M-m;2,m}Y_{4}^{M-m}(\mathbf{\hat{R}})\nonumber\\&
-\frac{2\xi}{3\pi^{\frac{3}{2}}R^{5}}\sqrt{\frac{5}{21}}C^{2,M}_{2,M-m;2,m}Y_{2}^{M-m}(\mathbf{\hat{R}})\nonumber\\&
+\frac{D_{5}}{R^{5}}Y_{0}^{0}(\mathbf{\hat{R}})\delta_{M,m},
\end{align}
\begin{equation}
c_{0,0}^{-5}(\vect R)=-\frac{6E_{5}}{\pi R^{5}}\sqrt{\frac{2\pi}{15}}Y_{2}^{M}(\mathbf{\hat{R}}),
\end{equation}

\begin{align}
c_{6,m}^{-6}(\vect R)=&\frac{1536b_{6}a^{2}}{5R^{6}}\sqrt{\frac{182}{\pi}}C^{2,M}_{8,M-m;6,m}Y_{8}^{M-m}(\mathbf{\hat{R}})\nonumber\\&
\frac{9984a_{6}a^{2}}{5R^{6}}\sqrt{\frac{7}{11\pi}}C^{2,M}_{6,M-m;6,m}Y_{6}^{M-m}(\mathbf{\hat{R}})\nonumber\\&
+\frac{5760a^{2}}{R^{6}}\sqrt{\frac{78\pi}{11}}C^{2,M}_{4,M-m;6,m}Y_{4}^{M-m}(\mathbf{\hat{R}}),
\end{align}
\begin{align}
c_{4,m}^{-6}(\vect R)=&-\frac{A_{6}}{\sqrt{35\pi}R^{6}}C^{2,M}_{2,M-m;4,m}Y_{2}^{M-m}(\mathbf{\hat{R}})\nonumber\\&
+\frac{B_{6}}{\pi R^{6}}\sqrt{\frac{2}{77\pi}}C^{2,M}_{4,M-m;4,m}Y_{4}^{M-m}(\mathbf{\hat{R}})\nonumber\\&
-\frac{C_{6}}{\pi R^{6}}\sqrt{\frac{2}{11\pi}}C^{2,M}_{6,M-m;4,m}Y_{6}^{M-m}(\mathbf{\hat{R}}),
\end{align}
\begin{align}
c_{2,m}^{-6}(\vect R)=&\frac{D_{6}}{5\sqrt{35}\pi^{\frac{3}{2}}R^{6}}C^{2,M}_{4,M-m;2,m}Y_{4}^{M-m}(\mathbf{\hat{R}})\nonumber\\&
+\frac{E_{6}}{3\pi^{\frac{3}{2}}R^{6}}C^{2,M}_{2,M-m;2,m}Y_{2}^{M-m}(\mathbf{\hat{R}})\nonumber\\&
+\frac{F_{6}}{2\sqrt{6}\pi^{\frac{3}{2}}R^{6}}Y_{0}^{0}(\mathbf{\hat{R}})\delta_{M,m},
\end{align}
\begin{equation}
c_{0,0}^{-6}(\vect R)=\frac{4G_{6}}{R^{6}}\sqrt{\frac{2\pi}{15}}Y_{2}^{M}(\mathbf{\hat{R}}),
\end{equation}
where
\begin{equation}
    w_{2}=21\sqrt{3}-10\pi,
\end{equation}
\begin{equation}
    b_{2}=9\sqrt{3}-4\pi,
\end{equation}
\begin{equation}
    c_{2}=9\sqrt{3}+4\pi,
\end{equation}
\begin{equation}
    d_{2}=87-16\sqrt{3}\pi,
\end{equation}
\begin{equation}
    w_{4}=16\pi+27\sqrt{3},
\end{equation}
\begin{equation}
    b_{4}=297\sqrt{3}-160\pi,
\end{equation}
\begin{equation}
    c_{4}=655\sqrt{3}\pi-3564,
\end{equation}
\begin{equation}
    d_{4}=9+\sqrt{3}\pi,
\end{equation}
\begin{equation}
    e_{4}=2\sqrt{3}\pi-9,
\end{equation}
\begin{equation}
    f_{4}=134\sqrt{3}\pi-729,
\end{equation}
\begin{equation}
    A_{1}=(792\sqrt{3}-865\pi+76\sqrt{3}\pi^{2})a+9\pi r_{s},
\end{equation}
\begin{equation}
    A_{3}=27\pi a a_{d}r_{d}+4b_{2}a^{3},
\end{equation}
\begin{equation}
    B_{3}=8(3429-1138\sqrt{3}\pi+280\pi^{2})a^{3}+5\sqrt{3}\pi a r_{d}a_{d},
\end{equation}
\begin{equation}
    A_{4}=16\left(-243\pi d_{2} r_{d}a_{d}+2 a f_{4}A_{1}\right)a^{2},
\end{equation}
\begin{equation}
    B_{4}=72\sqrt{3}\pi b_{2}a^{2}r_{d}a_{d}+16a^{3} e_{4}A_{1},
\end{equation}
\begin{equation}
    C_{4}=-9\sqrt{3}\pi c_{2}a^{2}r_{d}a_{d}+4d_{4}a^{3}A_{1},
\end{equation}
\begin{align}
    A_{5}=&3(27490131-9134582\sqrt{3}\pi+2245320\pi^{2})a^{3}\nonumber\\&
    +\frac{8505\sqrt{3}\pi}{2}aa_{g}r_{g},
\end{align}
\begin{equation}
    B_{5}=60b_{2}a^{3}+675aa_{g}r_{g},
\end{equation}
\begin{align}
    C_{5}=&\frac{1}{8\sqrt{7}\pi^{\frac{3}{2}}}\biggl[24(-127\sqrt{3}+70\pi)\xi+378\pi aa_{d}B_{3}r_{d}\nonumber\\&
    +315\sqrt{3}\pi^{2}a_{d}\left(240+a r_{d}'\right)\biggr],
\end{align}
\begin{equation}
    D_{5}=\frac{4\sqrt{\frac{2}{3}}\left(\xi+54b_{2}\pi a^{3}r_{d}a_{d} \right)}{3\pi^{\frac{3}{2}}},
\end{equation}
\begin{equation}
    E_{5}=10(28\pi-27\sqrt{3})a^{2}a_{d}+\frac{\pi\alpha}{3},
\end{equation}
\begin{equation}
    A_{6}=\frac{8(10\sqrt{3}\pi+81)A_{1}a^{3}}{\pi}-108\sqrt{3}\, w_{4}a^{2}a_{g}r_{g},
\end{equation}
\begin{equation}
    B_{6}=32(76\sqrt{3}-405)A_{1}a^{3}+288\sqrt{3}b_{4}w_{4}\pi a^{2}r_{g},
\end{equation}
\begin{equation}
    C_{6}=32(15020\sqrt{3}\pi-81729)A_{1}a^{3}+17280c_{4}w_{4}\pi a^{2}r_{g},
\end{equation}
\begin{align}
    D_{6}=&40\sqrt{5}\biggl[4f_{4}A_{1}a^{3}a_{d}r_{d}+108\pi(-8577+1600\sqrt{3}\pi)a a_{d}\nonumber\\&
    +24\pi d_{2}\eta-81\pi d_{2}a^{2}a_{d}\left(6r_{d}^{2}a_{d}+r_{d}'\right)\biggr],
\end{align}
\begin{align}
    E_{6}=&\sqrt{\frac{5}{7}}\biggl[16\sqrt{3}b_{2}\pi\eta+9\sqrt{3}b_{2}\pi (6r_{d}^{2}a_{d}+r_{d}')a^{2}a_{d}\nonumber\\&
    +12e_{4}A_{1}a^{3}a_{d}r_{d}+1944\pi(-129+20\sqrt{3}\pi)a a_{d}\biggr],
\end{align}
\begin{align}
    F_{6}=&-8\sqrt{3}\left[d_{4}a^{2}r_{d}A_{1}-9\pi (-621+100\sqrt{3}\pi)\right]aa_{d}\nonumber\\&
    +\pi c_{2}\left[-16\eta+9\left(6r_{d}^{2}a_{d}+r_{d}'\right)a^{2}a_{d}\right],
\end{align}
\begin{align}
    G_{6}=&-2D+\frac{495+36\sqrt{3}-70\pi}{25\pi^{2}}a E_{5}\nonumber\\&
    +\frac{4}{5\pi^{2}}(-17356761+8938062\sqrt{3}\pi-45240\pi^{2}\nonumber\\&
    +259200\sqrt{3}\pi^{3})a^{3}a_{d}
    +\frac{27(-81\sqrt{3}+40\pi)}{\pi}a a_{d}^{2}r_{d},
\end{align}
\begin{align}
    \alpha=&\frac{(-27\pi b_{2}r_{s}-2w_{2}A_{1})a^{5}r_{s}-180(27\sqrt{3}-28\pi)a^{2}a_{d}}{6\pi^{2}}\nonumber\\&
    -\frac{9b_{2}a^{4}r_{s}'-16b_{2}a\eta}{12\pi},
\end{align}
\begin{align}
    \eta=&-\frac{a}{144\pi^{2}}\biggl[54\pi a^{3}r_{s}A_{1}+81\pi^{2}(a^{2}r_{s}'+240a_{d})\nonumber\\&
    +(19\sqrt{3}\pi-88)\xi\biggr],
\end{align}
\begin{equation}
    \xi=-4a^{4}w_{2}A_{1}-54b_{2}\pi a^{4}r_{s}.
\end{equation}
\begin{equation}
    P_{2,0}^{(-1)}=\frac{16}{15}\sqrt{\frac{2}{5}}b_{2}a^{3},
\end{equation}
\begin{equation}
    P^{(-2)}_{2,0}=-\frac{16}{45}\sqrt{\frac{2}{5}}a^{3}A_{1},
\end{equation}
\begin{equation}
    P^{(-3)}_{2,0}=\frac{2}{45\pi}\sqrt{\frac{2}{5}}\xi,
\end{equation}
\begin{equation}
    P_{0,2}^{(-3)}=-81\sqrt{\frac{5}{2}}\pi a_{d},
\end{equation}
\begin{equation}
    P_{2,0}^{(-4)}=-\frac{64}{15}\sqrt{\frac{2}{5}}\eta,
\end{equation}
\begin{equation}
    P_{0,2}^{(-4)}=2016\sqrt{\frac{2}{5}}a a_{d},
\end{equation}
\begin{equation}
    P_{2,2}^{(-4)}=-\frac{2304aa_{d}}{5\sqrt{7}},
\end{equation}
\begin{equation}
    P_{4,2}^{(-4)}=\frac{3072a a_{d}}{35\sqrt{35}},
\end{equation}
\begin{equation}
    P_{2,0}^{(-5)}=-4\sqrt{\frac{2}{5}}\pi \alpha,
\end{equation}
\begin{equation}
    P_{0,2}^{(-5)}=-9\sqrt{10}c_{2}a^{2}a_{d},
\end{equation}
\begin{equation}
    P_{2,2}^{(-5)}=12\sqrt{7}b_{2}a^{2}a_{d},
\end{equation}
\begin{equation}
    P^{(-5)}_{4,2}=\frac{1296d_{2}a^{2}a_{d}}{\sqrt{105}},
\end{equation}
\begin{equation}
    P_{2,2}^{(-6)}=\frac{768a a_{d}}{5\sqrt{7}\pi}\left[4b_{2}a^{2}+27\pi r_{d}a_{d}\right],
\end{equation}
\begin{align}
    P^{(-6)}_{4,2}=\frac{768}{7\pi}\sqrt{\frac{3}{35}}aa^{d}& \biggl[8a^{2}(-3429+1138\sqrt{3}\pi-280\pi^{2})\nonumber\\&
    -5\sqrt{3}\pi r_{d}a_{d}\biggr],
\end{align}
\begin{equation}
    P_{2,0}^{(-6)}=\frac{192a}{5\pi}\sqrt{\frac{2}{5}}E_{5},
\end{equation}
\begin{equation}
    P_{0,2}^{(-6)}=-\frac{4608}{\sqrt{10}\pi}b_{2}a^{3}a_{d}.
\end{equation}
\begin{align}
    J_{1}=&-\frac{64 }{135 \pi ^3}\sqrt{\frac{2}{5}}(88 \sqrt{3}-57) \bigl(3040 \pi ^4 \sqrt{3}\nonumber\\&
    +225282 \pi ^2 \sqrt{3}+449064 \sqrt{3}-74272 \pi ^3\nonumber\\&
    -903879 \pi \bigr)\approx 371.103,
\end{align}
\begin{align}
    J_{2}=&\frac{32}{15 \pi ^2} \sqrt{\frac{2}{5}} \biggl[1520 \pi ^4 \sqrt{3}+4 \pi ^3 \left(228 \sqrt{3}-6719\right)\nonumber\\&
    +81 \left(437 \sqrt{3}-2816\right)+3 \pi  \left(27497 \sqrt{3}-43980\right)\nonumber\\&
    +\pi ^2 \left(54602 \sqrt{3}-27800\right)\biggr]\approx -176.377,
\end{align}
\begin{align}
    J_{3}&=\frac{96}{5 \pi }\sqrt{\frac{2}{5}} \left[-27 \sqrt{3}+32 \pi ^2+\pi  \left(12-69 \sqrt{3}\right)\right]\nonumber\\&
    \approx -265.527,
\end{align}
\begin{align}
    J_{4}=-\frac{96 }{5 \pi } \sqrt{\frac{2}{5}}b_2\approx3566.37,
\end{align}
\begin{equation}
\begin{split}
    J_{5}=-\frac{768 }{\pi }\sqrt{\frac{2}{5}} \left(81 \sqrt{3}-52 \pi \right)\approx-11.6812.
\end{split}
\end{equation}

\section{FUNCTIONS IN 111-EXPANSION\label{ref:function}}
\begin{equation}
    \gamma^{(0)}_{2,0}(\theta)=-\frac{5}{8}\sec\theta\left(2\cos\theta+3\cot\theta\csc\theta-3\theta\csc^{3}\theta\right),
\end{equation}
\begin{equation}
   \gamma_{2,0}^{(-1)}(\theta)=-\frac{5}{2}\left(3\theta\csc^{3}\theta+\sec\theta-3\csc^{2}\theta\sec\theta\right),
\end{equation}
\begin{equation}
   \gamma_{2,0}^{(-2)}(\theta)=\frac{15}{8}\csc^{3}\theta\sec\theta\left[2\theta(2+\cos 2\theta)-3\sin 2\theta\right],
\end{equation}
\begin{equation}
    \gamma^{(-3)}_{2,0}(\theta)=-\frac{5}{2}(3\theta\csc^{3}\theta+\sec\theta-3\csc^{2}\theta
    \sec\theta),
\end{equation}
\begin{equation}
    \gamma_{0,2}^{(-3)}(\theta)=\sec^{3}\theta,
\end{equation}
\begin{equation}
    \gamma_{2,0}^{(-4)}(\theta)=\frac{5}{8}\left[-2+3\csc^{2}\theta (-1+\theta\csc\theta\sec\theta)\right],
\end{equation}
\begin{equation}
    \gamma_{0,2}^{(-4)}(\theta)=\frac{1}{32}\csc\theta\sec^{3}\theta(12\theta+8\sin 2\theta+\sin 4\theta),
\end{equation}
\begin{align}
    \gamma_{2,2}^{(-4)}(\theta)=&\frac{5}{256}\csc^{3}\theta\sec^{3}\theta \nonumber\\&
    \times\left(-24\theta\cos 2\theta+9\sin 2\theta+\sin 6\theta\right),
\end{align}
\begin{align}
    \gamma_{4,2}^{(-4)}(\theta)=&\frac{105}{8192}\csc^{5}\theta\sec^{3}\theta\big[4\big(78\theta+96\theta\cos 2\theta\nonumber\\&
    +36\theta\cos 4\theta-36\sin 2\theta-29\sin 4\theta-4\sin 6\theta\big)\nonumber\\&
    +\sin 8\theta\big],
\end{align}
\begin{equation}
    \gamma_{2,0}^{(-5)}(\theta)=\sin\theta\tan\theta,
\end{equation}
\begin{equation}
 \gamma_{0,2}^{(-5)}(\theta)=\frac{1}{15}(8+6\cos 2\theta+\cos 4\theta)\sec^{3}\theta,
\end{equation}
\begin{equation}
   \gamma_{2,2}^{(-5)}=\frac{1}{7}(5+2\cos 2\theta)\tan^{2}\theta\sec\theta,
\end{equation}
\begin{equation}
    \gamma^{(-5)}_{4,2}(\theta)=\sin\theta\tan^{3}\theta,
\end{equation}
\begin{equation}
   \gamma_{2,2}^{(-6)}(\theta)=\frac{5}{128}\csc^{3}2\theta\left(24\theta-8
    \sin 4\theta+\sin 8\theta\right),
\end{equation}
\begin{align}
    \gamma_{4,2}^{(-6)}(\theta)=&\frac{21}{40960}\csc^{5}\theta\sec^{3}\theta\big(-480\theta-1200\theta\cos 2\theta \nonumber\\&
    +480\sin 2\theta+160\sin 4\theta+35\sin 6\theta\nonumber\\&
    -20\sin 8\theta+3\sin 10\theta\big).
\end{align}
%\begin{equation}
%    \overset{\cdot}{\gamma}_{2,0}^{(-6)}(\theta)=-\frac{1}{32} \theta  (5 \cos 2 \theta -4 \cos 4 \theta +\cos 6 \theta ) \csc ^3\theta  \sec \theta +\frac{1}{240} \left(-74 \cos 2 \theta +15 \csc ^2\theta +114\right) ,
%\end{equation}
%\begin{equation}
%    \overset{\cdot}{\gamma}_{0,2}^{(-6)}(\theta)=-\frac{1}{32} \theta  (5 \cos 2 \theta +4 \cos 4 \theta +\cos 6 \theta ) \csc \theta  \sec ^3\theta +\frac{1}{96} (12 \cos 2 \theta +\cos 4 \theta +17) \sec ^2 \theta ,
%\end{equation}

\section{BORN EXPANSION\label{app:Born}}

%Born approximation can be utilized when the interaction is weak. Wave function can be written in the following way:
%\begin{equation}
%    \Psi = \Psi_{0} + \hat{G}U\Psi_{0} + \hat{G}U\hat{G}U\Psi_{0}+\cdots.
%\end{equation}
%Here $\hat{G}=\frac{1}{-\hat{H_{0}}}$ is the Green operator, where $H_{0}$ is the kinetic operator of three bosons. The Green operator is defined in a six-dimensional space and the explicit form of Green function is:
%\begin{equation}
%    G(\mathbf{\rho},\mathbf{\rho'})=-\frac{\sqrt{3}}{8\pi^{3}}\frac{1}{\left[(\mathbf{R}-\mathbf{R'})^{2}+\frac{3}{4}(\mathbf{s}-\mathbf{s'})^{2}\right]^{2}},
%\end{equation}
%where $\mathbf{\rho}=(\mathbf{R},\mathbf{s})$ is the hyperradius.

To derive the Born approximation of $D$, we consider the $L=2$ and $M=2$ collision of the three bosons, for which
$\Psi_{0}=2(R_{x}+i R_{y})^{2}+\frac{3}{2}(s_{x}+i s_{y})^{2}$. % We assume that the interaction is short range (i.e. $V\left(s_{i}\right)=0$ if $s_{i}>r_{e}$ and $V_{3-body}\left(s_{1},s_{2},s_{3}\right)=0$ if $s_{1}>r_{e}$ or $s_{2}>r_{e}$ or $s_{3}>r_{e}$).
The first order term of the Born expansion is
\begin{equation}
\Psi_{1}=\hat{G}U\Psi_0=\Psi_{1}^{\text{3-body}}+\sum_{i=1}^3\Psi_{1}^{(i)}, \label{eq:born1}
\end{equation}
where $\Psi_1^\text{3-body}=\hat{G}V_\text{3-body}\Psi_0$,
\begin{widetext}
\begin{align}
    \Psi_{1}^{(i)}=\hat{G}V_i\Psi_0&=-\frac{M_B}{\hbar^{2}}\int d\mathbf{s'}d\mathbf{R'} \frac{1}{2\pi^{3}}\sqrt{\frac{2\pi}{5}}\frac{V(s')(2R'^{2}Y_{2}^{M}(\hat{\vect R}')+\frac{3}{2}s'^{2}Y_{2}^{M}(\hat{\vect s}'))}{\left[(\vect R_i-\mathbf{R'})^{2}+\frac{3}{4}(\vect s_i-\mathbf{s'})^{2}\right]^{2}}\nonumber\\&
    =-8\sqrt{\frac{2\pi}{15}}R_{i}^{2}Y_{2}^{M}(\hat{\vect R}_i)\left[a_{0}(s_i)+\frac{\alpha_{1}(s_i)}{s_i}\right]-\frac{6}{5}s_{i}^{2}Y_{2}^{M}(\hat{\vect s}_i)\left[a_{0}(s_i)+\frac{\alpha_{5}(s_i)}{s_{i}^{5}}\right],\label{eq:born1i}
\end{align}
and $V_i\equiv V(s_i)$.
In order to calculate the contribution of the three-body potential to $\Psi_1$, we can re-express the three-body potential as $V_\text{3-body}\left(s,R,\theta\right)$, where $\theta$ is the angle between $\mathbf{s}$ and $\mathbf{R}$. We find
\begin{equation}
    \Psi_{1}^{\text{3-body}}=-\frac{M_B}{2\pi^{3}\hbar^{2}}\sqrt{\frac{2\pi}{5}}\int d\mathbf{s'}d\mathbf{R'}\frac{V_{\mathrm{3-body}}(s',R',\theta)[2R'^{2}Y_{2}^{M}(\hat{\vect R}')+\frac{3}{2}s'^{2}Y_{2}^{M}(\hat{\vect s}')]}{\left[(\mathbf{R}-\mathbf{R'})^{2}+\frac{3}{4}(\mathbf{s}-\mathbf{s'})^{2}\right]^{2}}. \label{eq:3bodyBorn}
\end{equation}
\end{widetext}
Expanding \Eq{eq:3bodyBorn} at large $B$, we get
\begin{equation}
    \Psi_{1}^{\text{3-body}}\simeq -\frac{4\Delta}{ B^{8}}\sqrt{\frac{2\pi}{15}}\sum_{i=1}^3s_{i}^{2}Y_{2}^{M}(\mathbf{\hat{s}}_{i}).
\end{equation}

% The \nocite command causes all entries in a bibliography to be printed out
% whether or not they are actually referenced in the text. This is appropriate
% for the sample file to show the different styles of references, but authors
% most likely will not want to use it.
\nocite{*}

\bibliography{references}% Produces the bibliography via BibTeX.

\end{document}